\documentclass[sigconf]{acmart}

\usepackage{multirow}
\usepackage{hyperref}
\usepackage{amsfonts}       
\usepackage{amsmath,amssymb,amsthm}
\usepackage{subcaption}

\AtBeginDocument{%
  \providecommand\BibTeX{{%
    \normalfont B\kern-0.5em{\scshape i\kern-0.25em b}\kern-0.8em\TeX}}}

\setcopyright{acmcopyright}
\copyrightyear{2018}
\acmYear{2018}
\acmDOI{XXXXXXX.XXXXXXX}

\acmConference[Conference acronym 'XX]{Make sure to enter the correct
  conference title from your rights confirmation emai}{June 03--05,
  2018}{Woodstock, NY}
\acmPrice{15.00}
\acmISBN{978-1-4503-XXXX-X/18/06}

\begin{document}

\title{Split Two-tower Model for Efﬁcient and Privacy-Preserving Cross-device Federated Recommendation}

\author{Jiangcheng Qin}
\email{qjc@zjhu.edu.cn}
\affiliation{%
  \institution{Ningbo University, Huzhou University}
  \country{China}
}

\author{Baisong Liu}
\email{lbs@nbu.edu.cn}
\affiliation{%
    \institution{Ninbo University}
      \country{China}
}

\author{Xueyuan Zhang}
\email{zhangxueyuan.cap@gmail.com}
\affiliation{%
    \institution{Ninbo University}
      \country{China}
}
\author{Jiangbo Qian}
\email{qianjiangbo@nbu.edu.cn}
\affiliation{%
    \institution{Ninbo University}
      \country{China}
}

\renewcommand{\shortauthors}{Qin, et al.}

\begin{abstract}
  Federated Recommendation can mitigate the systematical privacy risks of traditional recommendation since it allows the model training and online inferring without centralized user data collection. Most existing works assume that all user devices are available and adequate to participate in the Federated Learning. However, in practice, the complex recommendation models designed for accurate prediction and massive item data cause a high computation and communication cost to the resource-constrained user device, resulting in poor performance or training failure. Therefore, how to effectively compress the computation and communication overhead to achieve efficient federated recommendations across ubiquitous mobile devices remains a significant challenge. This paper introduces split learning into the two-tower recommendation models and proposes STTFedRec, a privacy-preserving and efficient cross-device federated recommendation framework. STTFedRec a-chieves local computation reduction by splitting the training and computation of the item model from user devices to a performance-powered server. The server with the item model provides low-dimensional item embeddings instead of raw item data to the user devices for local training and online inferring, achieving server broadcast compression. The user devices only need to perform similarity calculations with cached user embeddings to achieve efficient online inferring. We also propose an obfuscated item request strategy and multi-party circular secret sharing chain to enhance the privacy protection of model training. The experiments conducted on two public datasets demonstrate that STTFedRec improves the average computation time and communication size of the baseline models by about 40$\times$ and 42$\times$ in the best case scenario with balanced recommendation accuracy. 
\end{abstract}




\maketitle

\section{Introduction}
Recently, privacy-preserved recommendation have become a research hotspot due to growing concerns about user privacy\cite{mothukuri2021}, data security\cite{chai2020secure}, and strict government regulations such as \emph{GDPR}\footnote{https://gdpr.eu/} and \emph{CPRA}\footnote{https://thecpra.org/}. \emph{Federated Learning} (FL)\cite{mcmahan2017communication} is considered one of the effective privacy-preserving machine learning paradigms\cite{kairouz2021advances}. Participants of FL collaborate to train a model under the coordination of a central server while keeping the training data decentralized. FL can mitigate the systematic privacy risks of traditional centralized machine learning, which fits the need for data security and privacy protection in recommendation systems (RS)\cite{yang2020federated}. Ammad-ud-din et al.\cite{ammad2019federated} proposed the \emph{Federated Collaborative Filtering} (FCF), the FCF validated that a collaborative filter can be federated without a loss of accuracy compared to centralized implementation. Following the work of FCF, Flanagan et al.\cite{flanagan2020federated} proposed the \emph{Federated Multi-view Matrix Factorization} (FED-MVMF). FED-MVMF leverages multiple user data sources to facilitate model training, enhance model accuracy, and handle cold start problems. For specific application scenarios, Chen et al.\cite{chen2020practical} proposed \emph{PriRec} for Federated Point-of-Interest recommendation. Qi et al.\cite{qi2020privacy} proposed \emph{FedNewsRec} for Federated news recommendation. And Zhou et al.\cite{zhou2019privacy} proposed a federated social recommendation framework with Big Data Support. Most existing works assume all user devices are available and adequate for participant FL training.

However, in practice, FL based recommendation still faces the following challenges. On the one hand, modern recommendation systems are designed for accurate prediction and constructed with large model sizes\cite{acun2021understanding}, making resource-limited mobile devices inefficient for local model training\cite{kairouz2021advances}\cite{anelli2019towards}. Many devices cannot meet the performance requirements of the FL, resulting in poor model accuracy or even training failure. On the other hand, industrial-grade recommendations have massive item data, connecting massive users with items from huge volumes, often in the millions to billions\cite{yang2021improvement}. The centralized recommendation can mitigate this part of the work under strict latency requirements with the recall mechanism. However, since the server cannot access the user's data, massive raw item data must be delivered to the mobile device, causing considerable communication and computation overhead. In summary, a significant challenge for FL based recommendations to be applied in real-world environments is,

\emph{How to perform \textbf{efficient} and \textbf{privacy-preserving} FL based recommendations across the \textbf{resource-limited} ubiquitous mobile device.}

Specifically, we focus on two compression objectives of practical value for FL based RS. (a) \textbf{Server broadcast compression}\cite{li2020federated} to reduces communication costs when devices are involved in training and online prediction. (b) \textbf{Local computation reduction}\cite{lim2020federated}\cite{wang2020convergence} to reduces the computational workload of the device.

\textbf{Choices of recommendation model.} The two-tower model is widely used in traditional recommendation systems to recall massive items\cite{yang2020mixed}\cite{wang2021cross}. A unique feature of the two-tower model is that feature modeling for users and items is two independent sub-networks and can be cached separately\cite{huang2013learning}. Only the similarity matching of user and item embeddings is needed when online prediction is performed. Therefore, it is feasible to split the two independent sub-networks, and their training and online inferring can be performed within different entities, which matches our compression objectives.

\textbf{Choices of privacy-enhancing techniques.} Currently, many privacy-enhancing techniques have been used in FL since the local gradients may leak private information\cite{kairouz2021advances}\cite{anelli2019towards}, such as \emph{Differential Privacy} (DP)\cite{dwork2014algorithmic}, \emph{Secure Multiparty Computation} (SMPC)\cite{lindell2005secure}, and \emph{Homomorphic Encryption} (HE)\cite{fang2021privacy}. Since DP introduces random noises to the gradients that affect model accuracy and HE will cause too much extra computation\cite{dwork2016calibrating}\cite{pulido2021privacy}, we choose SMPC, which can provide security guarantees on multi parties' gradient aggregation.

\textbf{Dealing with computation and communication overhead.} Although some methods have been proposed to improve the efficiency of FL based RS, they do not explicitly address the above issues. Muhammad et al.\cite{muhammad2020fedfast} proposed FedFast to accelerate the training efficiency by building user clusters and allowing parameter sharing. Khan et al.\cite{khan2021payload} introduce the multi-arm bandit mechanism to tackle the item-dependent payloads problem. However, these two methods cannot reduce the device's workload during the training phase. Qin et al.\cite{qin2021novel} proposed PPRSF that enables the server to use a portion of the users' explicit feedback to build a recall model, resulting privacy exposure. Different from their work, in this paper, we proposed a split learning based workload compression method that effectively compresses the device computation and communication load under privacy protection.

This paper proposes a \textbf{S}plit \textbf{T}wo-\textbf{T}ower \textbf{Fed}erated \textbf{Rec}ommenda-tion framework (STTFedRec). In our proposed approach, based on the unique structure of the two-tower model, we use split learning to separate the user model (user tower) and item model (item tower) and place them on the user device and server, respectively. The user device is only responsible for the computation of the user tower, achieving the goal of model compression. For online prediction, instead of complete raw item data, the server will provide the user device with low-dimensional item embeddings to achieve the goal of communication compression. Finally, the user device only needs to perform the similarity computing between the user and item embeddings for item ranking, retains the end-to-end computing benefits of a centralized recommendation system, and achieves the goal of computation efficiency optimization.

\textbf{Summary of experimental results.} We conduct experiments on two public benchmark dataset, i.e., MovieLens1M and BookCrossings. The results demonstrate that (1) STTFedRec achieves the same performance as existing FL based recommendation models, Split learning did not affect the performance of the FL two-tower model. (2) STTFedRec improves the average computation time and communication size of the baseline models by about 40$\times$ and 42$\times$ in the best case scenario. (3) The proposed multi-party circular secret sharing chain does not put additional strain on the computing and communication of user devices, and its communication volume is not affected by the number of participating members.

\textbf{Contributions.} We summarize our main contributions below: (1) We propose STTFedRec, a privacy-preservsed and efficient cross-device recommendation framework, which can significantly reducing the computation and communication overhead on resource-constrained mobile devices. (2) We propsoe an obfuscated item request strategy and multi-party circular secret sharing chain to further enhance privacy during the STTFedRec training phase. (3) We conduct thorough experiments on real-world datasets to verify the effectiveness and efficiency of STTFedRec.

\section{Recommendation Model and Tools}

This section present the selected recommendation model and related techniques to STTFedRec.

\subsection{The Two-tower Model}

The two-tower model originated from \emph{the Deep Semantic Similarity Model} (DSSM)\cite{huang2013learning}. Utilizing two-tower models have become a popular approach in several natural language tasks and have proven to be outstandingly efficient in building large-scale semantic matching systems. Since semantic matching itself is a ranking problem that coincides with the recommendation scenario, the two-tower model was naturally introduced into the recommendation domain and is widely used in advertising (Facebook)\cite{huang2020embedding}, information retrieval (Google)\cite{wang2021dcn}, and recommendation systems (Youtube)\cite{yi2019sampling}.

Formally, utilize the two-tower model to build a recommendation system, the goal is to retrieve a subset of items for a given user for subsequent ranking. Denoting users' feature vector as $\{x_i\}^N_{i=1}$, items' feature vector as $\{y_j\}^M_{j=1}$. Build two embedding functions, $u: X\times R^d \rightarrow R^k$, $v: Y\times R^d \rightarrow R^k$, mapping user and candidate item to a $k$-dimensional vector space. The output of the model is the inner product of the two embeddings by Eq.(1). The training object is to learn the parameters $\theta_u$ and $\theta_v$ based on the training set $\mathcal{D}:=\{(x_i,y_i,r_i)\}^T_{i=1}$, where the $r_i$ is user feedback, indicates whether the user clicked, watched or rated an item.

\begin{equation}
    Sim(x,y)=<u(x,\theta_u),v(y,\theta_v)> \label{pythagorean}
\end{equation}

The item retrieval problem can be considered as a multiclassification problem, given a user $X$, the probability of getting $y$ from $M$ items can be calculated using the softmax function as Eq.(2). The log-likelihood loss funcion as Eq.(3).

\begin{equation}
  P(y|x, \theta_u, \theta_v) = \frac{e^{sim(x,y)}}{\sum_{j=1}^M e^{sim(x,y_j)}}
\end{equation}
\begin{equation}
  L_T(\theta_u, \theta_v):=-\frac{1}{T}\sum_{i\in|T|}{r_ilog(P(y_i|x_i, \theta_u,\theta_v ))}
\end{equation}

The unique feature of the two-tower model is that the structure of the user and item representation are two independent sub-networks, as shown in Figure \ref{STTFedRec}(a). The two independent towers are cached separately, and only similarity operations are performed in memory during online prediction. It is worth noting that this enables the training and computation process of these two independent sub-networks to be placed separately on the user and server sides in a FL setting. Based on these properties and combined with our optimization objectives, we choose the DSSM model as the basic model of STTFedRec.

\subsection{Secure Multiparty Computing (SMPC)}

SMPC is a cryptographic tool which enables multiple participants to properly perform distributed computing tasks in a secure manner\cite{lindell2005secure}. Fomally, denote $n$ participants $\{\mathcal{P}_i \}_{i \in [1,n] }$, wish to jointly compute the $n$ element function $f(x_1,x_2,...,x_n)$ using their respective secret inputs $x_i$, and gets output $y_i$. SMPC protocols ensure, at the end of the protocol, each participant does not receive any additional information beyond its own inputs and outputs and the information that can be derived from them.

\begin{equation}
f(x_1,x_2,...,x_n)=(y_1,y_2,...,y_n)
\end{equation}

Here, we present the Addition SMPC protocols which we will use later in our framework for secure gradient aggregation. Take two secret shares $\tt Shr()$ as inputs from two parties, $\mathcal P \it _{i\in\{0,1\}}$ Locally calculate and return $[a]_i+[b]_i$.
\begin{equation}
  \tt Add(Shr([a]),Shr([b])) = [a]_i+[b]_i
\end{equation}

With Eq.(4) and Eq.(5), members participating in FL can hide their actual gradient and thus maintain privacy, while the server can achieve secure gradients aggregation.

\subsection{Split Learning}

Split learning\cite{gupta2018distributed} is a collaborative deep learning technique. In contrast to federal learning settings that focus on data partitioning and communication patterns, the critical idea behind split learning is to divide the execution of the model by layer between the client and the server for both the training and inference processes of the model\cite{singh2019detailed}.

\begin{figure}[h]
  \centering
  \includegraphics[width=0.8\linewidth]{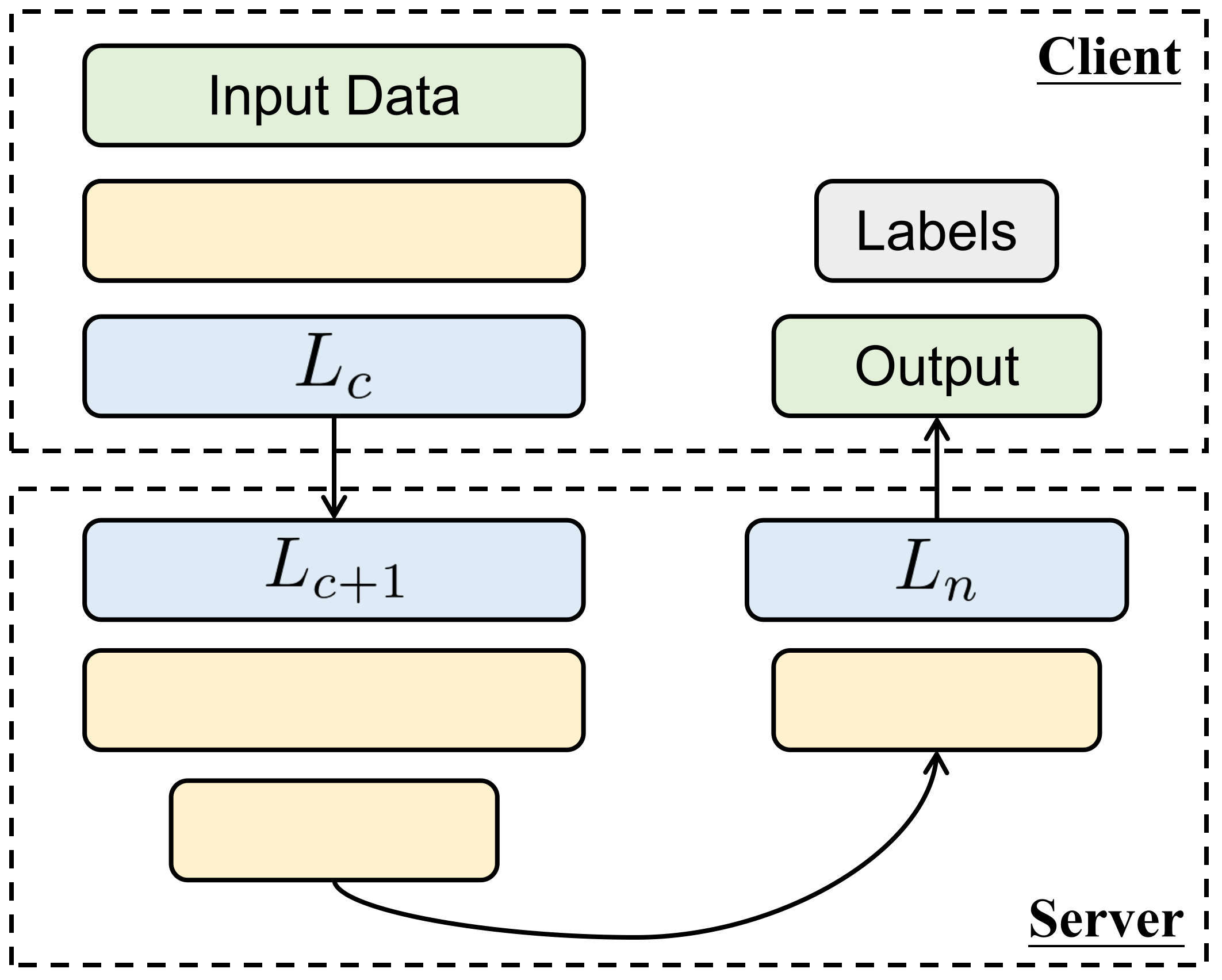}
  \caption{U-Shaped split learning configuration\cite{kairouz2021advances}}
  \Description{The U-shape split learning is privacy preserved variation since the raw data and labels are not transferred between the client and the server. }
  \label{splitlearning}
\end{figure}

The U-shaped split learning is privacy preserved variation since the raw data and labels are not transferred between the client and the server\cite{vepakomma2018split}. Formally, a client holds a deep learning task with $n$ computation layers $L_i, i\in[1,n]$. The client computes the forward propagation until it reaches the cut layer $L_c$ and $c<n$. The output of $L_c$ is then send to a computationally powerful server, which completes the forward propagation computation of $L_{c+1}\rightarrow L_n$. The output the $L_n$ will send back to the client and compute the loss function with labels. This completes a round of forward propagation. The gradients can then be backpropagated from the last layer until the cut layer in a reverse path. The cut layer's gradients will be sent back to the client, where the rest of the backpropagation is completed. The above process continues until the model converges. This setup is shown in Figure \ref{splitlearning}.

Based on split learning, we can extract part of the computation module of a recommendation model from resource-constrained mobile devices, compressing the computation overhead and offloading it to a server with superior performance. To the best of our knowledge, we propose the first FL based RS framework that uses split learning to reduce the computation and communication overhead of user devices.

\section{STTFedRec Framework}

This section describes the proposed STTFedRec Framework in detail. We first present the problem definition, and then present the overall framework, and training and online prediction process separately. We conclude with a discussion of the scalability of STTFedRec and possible variants.

\subsection{Problem Definition}
We consider a FL based recommendation problems, where we have a set of users  $\mathcal{U}=\{u_1,u_2,...\}$  and items $\mathcal{V}=\{v_1, v_2,...\}$ . Given a user $u$, his private data is denote as $\mathcal{D}_u:=\{(x_u,v_j,r_j)\}^M_{j=1}$ are locally stored on his own device, $x_u$ is user's personalized features and $r_j$ is the data label. Here $x_u\in \mathcal{X}$ and $y_j\in \mathcal{Y}$ are both mixtures of a wide variety of features (e.g., sparse IDs and dense features) .

We aim to build a two-tower model with two different parameterized embedding functions, including an user model $\mathcal{M}_u(x,\Theta _u)$ and an item model $\mathcal{M}_v(v,{\Theta}_v)$. The final output of the two-tower model is the similarity matching of user embeddings and item embeddings to provide a fast recall of the set of candidate items to the user from a large number of items.

The goal is to learn model parameters $\Theta_u$ and $\Theta_v$ collaborately from the training dataset $\mathcal D$ distributed in each user device, reducing computation and communication across participating devices and maintaining data privacy.

\subsection{Framework Overview}

\begin{figure*}[t]
  \centering
  \includegraphics[width=0.8\textwidth]{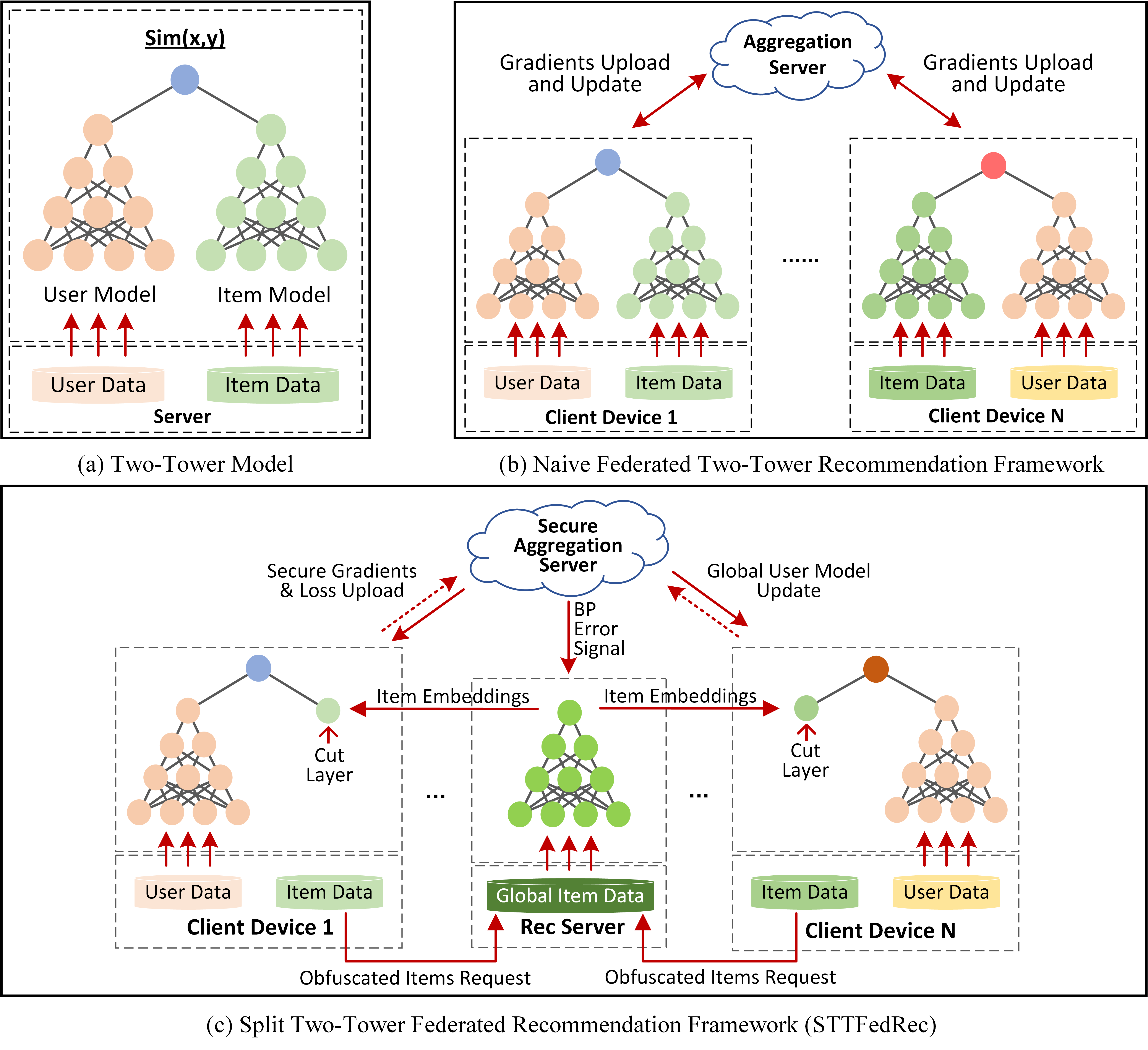}
  \caption{Illustration of (a)original two-tower model, (b)naive federated two-tower recomendation framework, and (c)split two-tower federated recommendation framework (STTFedRec)}
  \label{STTFedRec}
  \Description{the figure (a) illustrate the original structure of two-tower model, the figure (b) illustrate the naive setting of federated two-tower model that client device responsible for the training of entire model, the figure (c) illustrate the framework structure of STTFedRec and split the item model into a server, and client device is only responsible for the training of user model}
\end{figure*}

The framework of STTFedRec is shown in Figure \ref{STTFedRec}(c). There are $N$ different user devices as clients $\mathcal{U}$ participate STTFedRec with their mobile devices. The client's device maintains the local user profile $x$ and interaction data $r$, while the recommendation server maintains the massive global item data $\mathcal{V}$. Moreover,

\textbf{Towards the goal of computation and communication compression.} Unlike the naive setting of the federated two-tower model, illustrated in Figure \ref{STTFedRec}(b), we split the user model and item model of independent sub-networks based on the split learning. The user device is only responsible for the computing of the user tower, which compresses the computation overhead on a client device. The powerful recommendation server is responsible for the training and computation of item tower and providing the item embeddings instead of complete raw item data to the client device, which reduces the communication overhead of the user devices. When online inferring, the user device only needs to perform the similarity matching with cached user embeddings and items embeddings, retains the end-to-end computing benefits of a centralized recommendation system, and achieves computation efficiency optimization.

\textbf{Towards the goal of privacy-preserving.} STTFedRec leverages FL to allow multiple user devices to participate in model training simultaneously without exposing their local raw data. When user devices request item embeddings from the server, obfuscated item requests are generated by random negative sampling so that the server cannot obtain the actual labels of user interaction. We also employ a secure aggregation server (SAS) instructing participants to upload local gradient and loss via SMPC protocol (Section 2.2). SAS performs privacy-preserved global user model update by secure aggregated local gradients, and the secure aggragated loss will send to the recommendation server as backpropagation (BP) error signal for global item model update. The setting is secure against semi-honest adversary from restoring user data through the fully exposed local gradients and loss.

\subsection{Training and Online infering}

We here present the details of STTFedRec training process, including following six steps:

\textbf{Step 1.Initialization.} A recommendation server initialization of training process by defines the model structure of $\mathcal{M}_u(x,\Theta_u)$ and $\mathcal{M}_v(v, \Theta_v)$, and initializes the model parameters as $\Theta_u^0$ and $\Theta_v^0$. Randomly select a subset of available clients $\mathcal{C}_k^0 = \{c_k\}^K_{k=1}$ for the first round of training, where $\mathcal{C}_k^0\in \mathcal{U}$ and the group size denote as $K= |\mathcal{C}_k^0|$. Send the initialized $\mathcal{M}_u$ and $\Theta_u^0$ to each client in the subset of $\mathcal{C}_k^0$ and aggregation server. The superscript of each parameter indicates the current training round.

\textbf{Step 2.Obfuscated item request.} The client gets the item embeddings by initiating an obfuscated item request to the recommendation server. At training round $t$, the client $c_k^t$ randomly select a subset of non-clicked items' ID as $\mathcal{V}_k^-$ and a subset of clicked items' ID as $\mathcal{V}_k^+$, and the two items' ID sets form the obfuscated item request $\mathcal{V}_k^t = \{v^t_{kj}\}^T_{j=1}$, where the $T = |\mathcal{V}_k^t|$ as number of items requested. The proportion of clicked items denoted as $\rho = {|\mathcal{V}^-_k|}/{|\mathcal{V}^t|}$ is the negative sampling rate. The server will return items embeddings $\mathcal{I}_k^t = \{i^t_{kj}\}^T_{j=1}$ compute by Eq.(6)

\begin{equation}
  \mathcal{I}_k^t = \mathcal{M}_v(\mathcal{V}_k^t, \Theta_v^t) \\
\end{equation}

\textbf{Step 3.Client local training.} At training round $t$, a client $c_k^t$ perform local training to obtain the training loss and gradients based on the batch training set $\mathcal{B}^t_k = \{(x_k, v^t_{kj}r_j)\}_{j=1}^T$.
The user embeddings $u_k^t$ is compute locally by Eq.(7). We obtain the average loss $L_{\mathcal{B}_k}^t$ of the batch training set by Eq.(8). The gradients of local user model denote as $g^t_{ku}$ and compute by Eq.(9).

\begin{equation}
  u^t_k = \mathcal{M}_u(x_k, \Theta^t_u)
\end{equation}

\begin{equation}
  \begin{aligned}
    sim(x_k, v^t_{kj}) = <u^t_k, i_{kj}^t> = \frac{u^t_k \cdot i_{kj}^t}{||u^t_k||||i_{kj}^t||} \\
    P(v^t_{kj}|x_k,\Theta_u^t,\Theta_v^t) = \frac{e^{sim(x_k, v^t_{kj})}}{\sum_{j=1}^T e^{sim(v^t_{kj})}} \\
    L_{\mathcal{B}_k}^t = -\frac{1}{T}\sum^T_{j=1}r_j\cdot log(P(v^t_{kj}|x_k,\Theta_u^t,\Theta_v^t))
  \end{aligned}
\end{equation}

\begin{equation}
  g^t_{ku} = \frac{\partial L_{\mathcal{B}_k^t}}{\partial \Theta_u^t}
\end{equation}

\begin{figure}[bt]
  \centering
  \includegraphics[width=\linewidth]{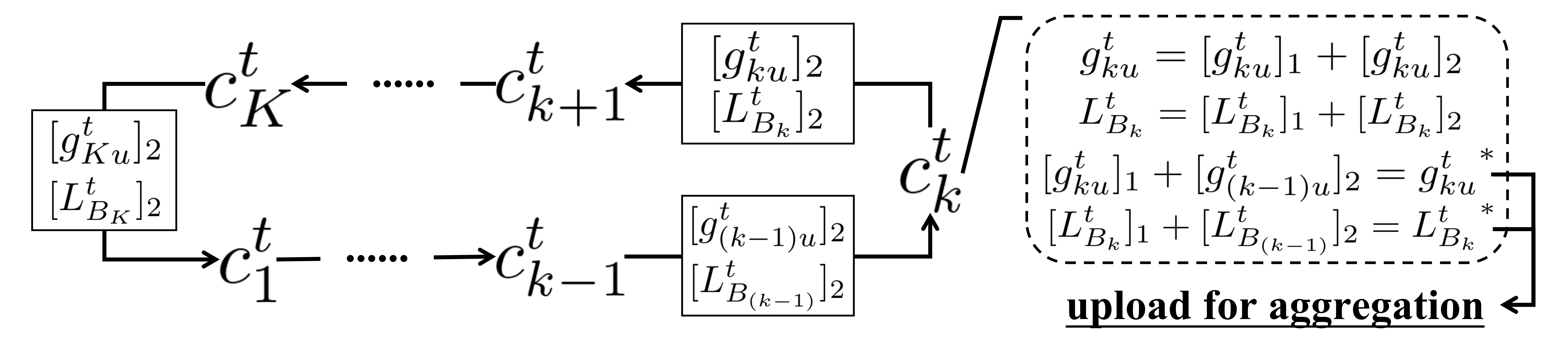}
  \caption{Illustration of secret sharing in the form of circular message delivery chain among clients for privacy-preserved local gradient and loss uploading}
  \label{figure3}
  \Description{}
\end{figure}

\textbf{Step 4.Gradients and loss secure aggregation.} At training round $t$, under the coordination of aggregation server, each client $c_k^t$ generates two random gradient matrics $[g_{ku}^t]_1$ and $[g_{ku}^t]_2$, and two random number $[L^t_{\mathcal{B}_k}]_1$ and $[L^t_{\mathcal{B}_k}]_2$ such that they satisfy Eq.(10). Delivers $[g_{ku}^t]_2$ and $[L^t_{\mathcal{B}_k}]_2$ to the next client $c_{k+1}^t$ that pass backwards one by one, and the last client $c_K^t$ will delivers $[g_{Ku}^t]_2$ and $[L^t_{\mathcal{B}_K}]_2$ to the first client $c_0^t$, forming a circular message delivery chain as illustrated in Figure \ref{figure3}. Sum the received gradient matrix and loss with reserved $[g_{ku}^t]_1$ and $[L^t_{\mathcal{B}_k}]_1$ to obtain a local mixed-gradients ${g_{ku}^t}^*$ and mixed-loss ${L^t_{\mathcal{B}_k}}^*$ by Eq.(11), and upload them to the aggregation server. The aggregation server computes the user model's average gradients $\overline{g_u^t}$ and loss $\overline{L^t}$ by Eq.(12). The average gradients will reserve for global user model update and the average loss will transfer to the recommendation server as BP error signal for global item model update.

\begin{equation}
  \begin{aligned}
  g^t_{ku} = [g_{ku}^t]_1+[g_{ku}^t]_2,\quad L^t_{\mathcal{B}_k} = [L^t_{\mathcal{B}_k}]_1+[L^t_{\mathcal{B}_k}]_2
  \end{aligned}
\end{equation}

\begin{equation}
  \begin{aligned}
    {g^t_{ku}}^*=[g_{ku}^t]_1+[g_{(k-1)u}^t]_2,\quad {L^t_{\mathcal{B}_k}}^*=[L^t_{\mathcal{B}_k}]_1+[L^t_{\mathcal{B}_{(k-1)}}]_2
  \end{aligned}
\end{equation}

\begin{equation}
  \begin{aligned}
    \overline{g_u^t} = \frac{1}{K}\sum_{k=1}^K {g_{ku}^t}^*,\quad \overline{L^t} = \frac{1}{K}\sum_{k=1}^k{L^t_{\mathcal{B}_k}}^*
  \end{aligned}
\end{equation}

\textbf{Step 5.Global user model update.} At training round $t$, the aggregation server performs the global user model update by FedAdam optimizer\cite{ReddiCZGRKKM21}, compute as Eq.(13), where the $\beta_1$, $\beta_2$, and $\tau$ are FedAdam's hyperparameters. The aggregation server will broadcast the updated global user model parameter $\Theta_u^{t+1}$ to all client devices for the next round of training or online inferring.
\begin{equation}
  \begin{aligned}
    m_u^t = \beta_1 m_u^{t-1}+(1-\beta_1)\overline{g_u^t}
    \\\upsilon_u^t = \beta_2\upsilon^t_{u-1}+(1-\beta_2)(\overline{g_u^t})^2
    \\\Theta_u^{t+1} = \Theta_u^t + \eta\frac{m_u^t}{\sqrt{\upsilon_u^t}+\tau}
  \end{aligned}
\end{equation}

\textbf{Step 6.Global item model update.} At training round $t$, the recommendation server receives the average BP error signals $\overline{L^t}$ from the aggregation server. The average gradients $\overline{g_v^t}$ is compute by Eq.(14). We also use the FedAdam optimizer to update the global item model, which compute as Eq.(15). The updated global item model $\mathcal{M}_v(y,\Theta_v^{t+1})$ will serve the next round obfuscated item request or online inferring. $t=t+1$, randomly select a new subset of available clients $\mathcal{C}^{t+1}_k$ and repeat the Step (2\textasciitilde6) for a new training round.

\begin{equation}
  \overline{g^t_v} =\frac{\partial\overline{L^t}}{\partial \Theta^t_v}
\end{equation}

\begin{equation}
  \begin{aligned}
    m_v^t = \beta_1 m_v^{t-1}+(1-\beta_1)\overline{g_v^t}
    \\\upsilon_v^t = \beta_2\upsilon^t_{v-1}+(1-\beta_2)(\overline{g_v^t})^2
    \\\Theta_v^{t+1} = \Theta_v^t + \eta\frac{m_v^t}{\sqrt{\upsilon_v^t}+\tau}
  \end{aligned}
\end{equation}

\textbf{Online inferring.} The recommendation server will compute the embeddings of massive item data based on the current global item model. The item embeddings $\mathcal{I}$ will be transmitted to each user device. The user can complete the online inferring by computing the inner product of $\mathcal I$ and locally cached user embeddings $u_k$. Compared to transferring the complete raw item data to the user and having the user device compute the corresponding embeddings, STTFedRec can significantly reduce the size of communication and computation overhead.

\subsection{Model Variations}
A three-layer, fully connected neural network is built for user and item feature representation in the most straightforward setup of the two-tower model, i.e., DSSM. The activation function of each hidden and output layer is tanh. Subsequent studies introduce convolutional neural networks\cite{shen2014latent}, recurrent neural networks\cite{zhai2016deepintent}, and attention mechanisms\cite{sun2021dsmn} in the representation layer to enhance the semantic feature representation. The unique structure of the two-tower model allows it to implement different feature engineering and mapping functions to adapt to different recommendation scenarios\cite{wang2021cross}. Elkahky et al.\cite{elkahky2015multi} use the multi-view DNN model to build a recommendation system that combines rich features from multiple domains to enhance the personalized expression of users. Huang et al.\cite{huang2020federated} verified the effectiveness of the Multi-view DSSM under FL setting. Google used the two-tower model in a large-scale video streaming recommendation system to model the interaction between user-item pairs and enhance user representation through user's video click sequences\cite{yi2019sampling}. Facebook also applies the two-tower structure in a large-scale social information retrieval model\cite{huang2020embedding}, and verifies its effectiveness in social recommendation scenarios.

This paper focuses on the computation and communication compression of STTFedRec for the user device in the FL setting. In the subsequent experiments, we select DSSM and CLSM to build STTFedRec-DSSM and STTFedRec-CLSM for performance and efficiency evaluation.

\section{Experiments}
This section performs extensive experiments on two public benchmark datsets to evaluate the proposed STTFedRec Framework. The experiments intend to answer the following research questions: \textbf{RQ1:} How does STTFedRec perform compare with baseline models? \textbf{RQ2:} Is the computation and communication overhead of STTFedRec significantly reduced compared to baseline models? \textbf{RQ3:} How does the hyperparameters user number $K$ and negative sampling rate $\rho$ effect the performance and efficiency of STTFedRec?

\subsection{Dataset and Experiment Settings}

We choose two public datasets to evaluate the performance of our proposed model, i.e., \emph{MovieLens-1M}\footnote{https://grouplens.org/datasets/movielens/}\cite{harper2015movielens} and \emph{Adressa-1week}\footnote{https://reclab.idi.ntnu.no/dataset/}\cite{gulla2017adressa}. The two datasets have significant difference in the size of the raw item data, which facilitates us to observe the impact of computation efficiency and communication cost of user devices. The MovieLens dataset contains about 1 million explicit ratings for movies. We treat explicit records with a rating greater than or equal to 1 as positive implicit feedback. The Adressa is a real-world online news dataset from a Norwegian news portal. Following\cite{hu2020graph}, we select the features include user\_id, news\_id, the title, and news profile and remove the stopwords in the news content. We also remove the users with less than 20 clicks to ensure a sufficient local user sample to participate in the FL training. Adressa has no user features, and we use five news titles that users have clicked as the user feature. For each user's clicked item, randomly sampled ten movies/news that users had not rated as negative feedback to generate obfuscated item requests. For both user and item features, we pre-coded with letter-tri-grams into fix-sized input vectors. We summarize the statistics of both datasets in Table \ref{Dataset}.

\begin{table}[ht]
  \caption{Overview of the datasets used in the experiment.}
  \label{Dataset}
  \centering
  \resizebox{\linewidth}{!}{
  \begin{tabular}{lcccccc}
    \toprule
    Dataset     & \# Users & \# Items & \# Clicks & Density(\%) & Item Data Size \\
    \midrule
    MovieLens-1M   & 6,022   & 3,043     & 995,154   & 5.4\%   & 171 Kbytes  \\
    Adressa-1week  & 186,255   & 14,732  & 2,103,852   & 0.07\%  & 1.4 Gigabyte  \\
    \bottomrule
  \end{tabular}
  }
\end{table}

\textbf{Hyperparameters.} Adam optimizer with an initial learning rate $10^{-4}$ with decay factor of $0.01$ after every $10$ rounds of training, and set $\beta_1=0.9$ and $\beta_2=0.99$. The select user number $K$ for a single round of training is searched in $\{25, 50, 100, 200\}$ and set as $50$ for MovieLens and $150$ for Adressa. The local batch size $T$ is set as $10$. $\rho$ is searched in $\{0.9, 0.8, 0.7, 0.6, 0.5\}$ and set as $0.9$. The dropout rate is set to 0.2. We initialize weights and bias parameters by a Gaussian distribution with a mean of $0$ and a standard deviation of $0.1$.
The hyperparameters of FCF and FedMVMF are consistent with Flanagan et al.\cite{flanagan2020federated}. For CLSM, the convolution window sizes is set to $3$. The models are written in Python3.7 programming environments by Pytorch. To simulate the computing power of mobile devices, we used a Macbook Pro with an M1 chip to simulate user devices and serialize the training process of each client for efficiency evaluation. A server equipped with dual-NVIDIA 2080Ti GPU was used for performance evaluation.

\textbf{Metrics.} We verify the performance of STTFedRec and baseline model in Top-K recommendations, adopt AUC, Precision@10, NDCG@10 as evaluation metrics. The AUC evaluates the overall accuracy while the Precision@10 and NDCG@10 concentrate on the very top of recommendation list. For efficiency, we report the training time (in millisecond), online infering time (in millisecond), and the communication size between a device and server (in megabytes). We repeat each experiment 5 times independently, and report average result.

\begin{table*}[h]
  \caption{Performance comparison results of different models in terms of AUC, Precision@10, and nDCG@10 on MovieLens and Adressa datasets. The values denote the mean $\pm$ standard deviation of metric values across 3 different model builds. }
  \label{table2}
  \centering
    \begin{tabular}{cccc|ccc}
    \toprule
    \multirow{2}*{Models}  & \multicolumn{3}{c|}{MovieLens}    & \multicolumn{3}{c}{Adressa}    \\

                      & AUC            & Precision@10   & nDCG@10        & AUC            & Precision@10   & nDCG@10    \\ \midrule
    DSSM              & 78.55$\pm$0.01 & 36.22$\pm$0.01 & 34.06$\pm$0.38 & 68.61$\pm$0.18 & 34.11$\pm$0.41 & 34.56$\pm$1.20 \\
    CLSM              & 80.11$\pm$0.09 & 38.20$\pm$0.08 & 35.27$\pm$1.11 & 70.39$\pm$0.31 & 35.14$\pm$0.32 & 35.86$\pm$2.14 \\ \midrule
    FCF               & 73.56$\pm$0.24 & 30.15$\pm$1.08 & 25.85$\pm$0.81 & 54.49$\pm$0.03 & 25.33$\pm$0.06 & 23.52$\pm$1.72 \\
    FedMVMF           & 74.75$\pm$0.03 & 32.07$\pm$0.88 & 26.99$\pm$1.31 & 60.99$\pm$0.01 & 28.28$\pm$0.10 & 25.91$\pm$0.58 \\
    FL-DSSM           & 69.34$\pm$0.82 & 27.66$\pm$1.42 & 22.86$\pm$1.47 & 66.37$\pm$0.07 & 31.20$\pm$0.22 & 29.19$\pm$0.94 \\
    FL-CLSM           & 73.53$\pm$1.25 & 31.17$\pm$1.68 & 25.33$\pm$0.59 & 68.99$\pm$0.11 & 33.17$\pm$0.09 & 30.36$\pm$1.68 \\ \midrule
    STTFedRec-DSSM    & 69.05$\pm$0.51 & 27.04$\pm$0.73 & 22.61$\pm$1.61 & 66.54$\pm$0.25 & 31.23$\pm$0.03 & 29.22$\pm$1.78 \\
    STTFedRec-CLSM    & 73.50$\pm$0.33 & 31.15$\pm$0.28 & 25.37$\pm$1.29 & 69.10$\pm$0.31 & 33.25$\pm$0.55 & 30.63$\pm$1.08 \\ \bottomrule
    \end{tabular}
\end{table*}

\subsection{Performance Comparison with Baselines (RQ1)}

We compare STTFedRec with several baseline methods, including the two-tower models in centralized version and in the naive FL setting. The baselines are listed as follows:
\begin{enumerate}
  \item \textbf{Centralized DSSM}\cite{huang2013learning}, the original version of two-tower model for information retrieve. In our experiments, we apply three-layer fully connected neural network for both towers, the distribution of neuron density is $<256, 128, 128>$, and the activation function for each layer is set as $tanh$.
  \item \textbf{Centralized CLSM}\cite{shen2014latent}, the convolutional-pooling structured two-tower model for item retrieval. In our experiments, we apply CLSM with one Convolutionl layer, one Max-pooling layer, and one fully connected layer as output layer, the dimension of max pooling layer set to $300$ and the output dimension is set to $128$.
  \item \textbf{FCF}\cite{ammad2019federated}, the FL based collaborative filtering recommendation model.
  \item \textbf{FedMVMF}\cite{flanagan2020federated}, the FL based multi-view matrix factorization recommendation model that also utilized side-information.
  \item \textbf{FL-DSSM}\cite{huang2020federated}, the naive setting of FL based DSSM for recommendation, the client device is responsible for training both user and item models, we apply same model structure as the centralized version.
  \item \textbf{FL-CLSM}, the naive setting of FL based CLSM for recommendation, its training approach is following Huang et al.\cite{huang2020federated}, the client device is responsible for training both user and item towers, we apply the same model structure as the centralized version.
  \item \textbf{STTFedRec-DSSM}, the FL based DSSM that implemented based on our proposed STTFedRec framework.
  \item \textbf{STTFedRec-CLSM}, the FL based CLSM that implemented based on our proposed STTFedRec framework.
\end{enumerate}

The performance comparison results are shown in Table \ref{table2}. From the Table, we have following observations: (1) compare to the centralized baseline model, the effect of applying STTMFedRec is within the performance loss of FL. The impact comes mainly from the bias introduced by gradient averaging and average loss based backpropogation. (2) Compare to the FL-DSSM and FL-CLSM under the naive FL setting, there is no performance loss in STTFedRec, i.e., using split learning and secure aggregation does not affect the performance of federated two-tower model. (3) Compare to the FCF and FedMVMF, STTFedRec is more suitable for scenarios with rich semantic information. Also, using interacted item data in user features and optimizing the representation layer can improve the performance of STTFedRec.

\begin{table*}[h]
  \caption{Efficiency comparison result on computation and communication cost of per round training and online inferring. The running time (in millisecond) and communication size (in megabytes) on MovieLens and Adressa datasets. }
  \label{table3}
  \centering

  \begin{tabular}{ccccc|cccc}
    \toprule
    \multirow{2}{*}{}     & \multicolumn{4}{c|}{MovieLens}  & \multicolumn{4}{c}{Adressa}        \\
                          & \multirow{2}{*}{FL-DSSM} & STTFedRec & \multirow{2}{*}{FL-CLSM} &STTFedRec & \multirow{2}{*}{FL-DSSM} & STTFedRec& \multirow{2}{*}{FL-CLSM} & STTFedRec\\
                          &         &DSSM       &         &CLSM      &          &DSSM    &          &CLSM  \\
    \midrule
    Device Training     & 1170ms     & 228ms     &4800ms    & 889ms     & 2570ms    & 447ms     &10930ms     &1560ms     \\
    Inferring Time      & 4090ms     & 55ms      &23860ms    & 55ms      & 19120ms   & 250ms     &79100ms    &250ms       \\
    Total Time          & 5260ms     & 283ms     &28660ms    & 944ms     & 21690ms   & 697ms     &90030ms    &1810ms\\
    <Improvement>       &\multicolumn{2}{c}{18$\times$}  &\multicolumn{2}{c|}{28$\times$} &\multicolumn{2}{c}{31$\times$} &\multicolumn{2}{c}{50$\times$}\\
    \\
    Training Comm.      & 5.42MB     & 2.71MB    & 4.70MB    & 2.35MB    & 8.68MB    & 4.34MB    & 17.04MB      &8.52MB       \\
    Inferring Comm.     & 0.14MB     & 0.05MB    & 0.14MB    & 0.05MB    & 557.80MB   & 7.19MB   &556.80MB      &7.19MB\\
    Total Comm.         & 5.56MB    & 2.76MB     & 4.84MB    & 2.40MB    & 566.48MB   & 11.53MB  &573.84MB      &15.71MB    \\
    <Improvement>       &\multicolumn{2}{c}{2$\times$} &\multicolumn{2}{c|}{2$\times$} &\multicolumn{2}{c}{49$\times$} &\multicolumn{2}{c}{36$\times$}\\
    \bottomrule
  \end{tabular}
\end{table*}

\subsection{Efficiency Comparison (RQ2)}

For each client $u$, the computational complexity of FL-DSSM for a single round of training is $Time\sim O((|x|+|v|\times T)\times |C_0| + \sum^D_{l=2}|C_{l-1}|\times |C_l|\times (1+T))$, where the $|C_l|$ denote the density of each computing layer, $|x|$ denote the input size of user features, and $|v|$ denote the input size of item features. The computational complexity of STTFedRec-DSSM is $Time\sim O(|x|\times|C_0|+\sum^D_{l=2}|C_{l-1}|\times|C_l|$ and the batch size of item model computation is left to the RS server. 

For the communication complexity, there are two communications between each client and the server at each training iteration, gradients upload and model download. For FL-DSSM, the communication complexity is $Space\sim O(2\times[(|x|+|v|)\times |C_0|+2\times\sum_{l=1}^D|C_{l-1}|\times |C_l|])$. The communication complexity of STTFedRec-DSSM is $Space\sim O(2\times(|x|\times|C_0| + \sum^D_{l=2}|C_{l-1}|\times|C_l|))$, user devices is no longer need to download and upload the parameters of the item model. Another important communication comes from the online inferring phase. The clients of FL-DSSM need to download the complete raw item data, while the clients of STTFedRec only need to download the processed item embeddings computed by the RS server.

We demonstrated the above theory in our experiment. We report the training time, inferring time, and communication size of STTFedRec and naive setting of FL based DSSM, CLSM in Table \ref{table3}. The Table shows that on the MovieLens dataset, STTFedRec improves average training time by $23\times$ and average communication overhead by $2\times$. On the Adressa dataset, STTFedRec improves average training time by $40\times$ and average communication overhead by $42\times$. More specifically, we observed that (1) the total communication cost and computation time of both STTFedRec and the naive setting of FL increase with the size of datasets and item feature size. Still, the increase rate of STTFedRec is slower than that of the naive setting. (2) In the case of a large and feature-dense dataset, STTFedRec has excellent benefits in reducing the computation and communication overhead. The original item data is replaced by the low-dimensional item feature embeddings provided by the server. Only similarity computation is needed to perform online inferring, which also validates the efficiency of STTFedRec for massive item recommendation scenarios. (3)STTFedRec shows better computational efficiency optimization when the computational complexity of the representation model of users and items is increased. The overall efficiency comparison result demonstrates that our proposed STTFedRec has better scalability than the naive setting of FL in terms of both running time and communication size. 

\subsection{Impact of Hyperparameters (RQ3)}
We further investigated the effect of hyperparameter settings ($\rho$ and $K$) on the performance and efficiency of STTFedRec on the Adressa dataset.

\begin{figure}[h]
\centering
    \begin{subfigure}{0.49\linewidth}
		\centering
		\includegraphics[width=\linewidth]{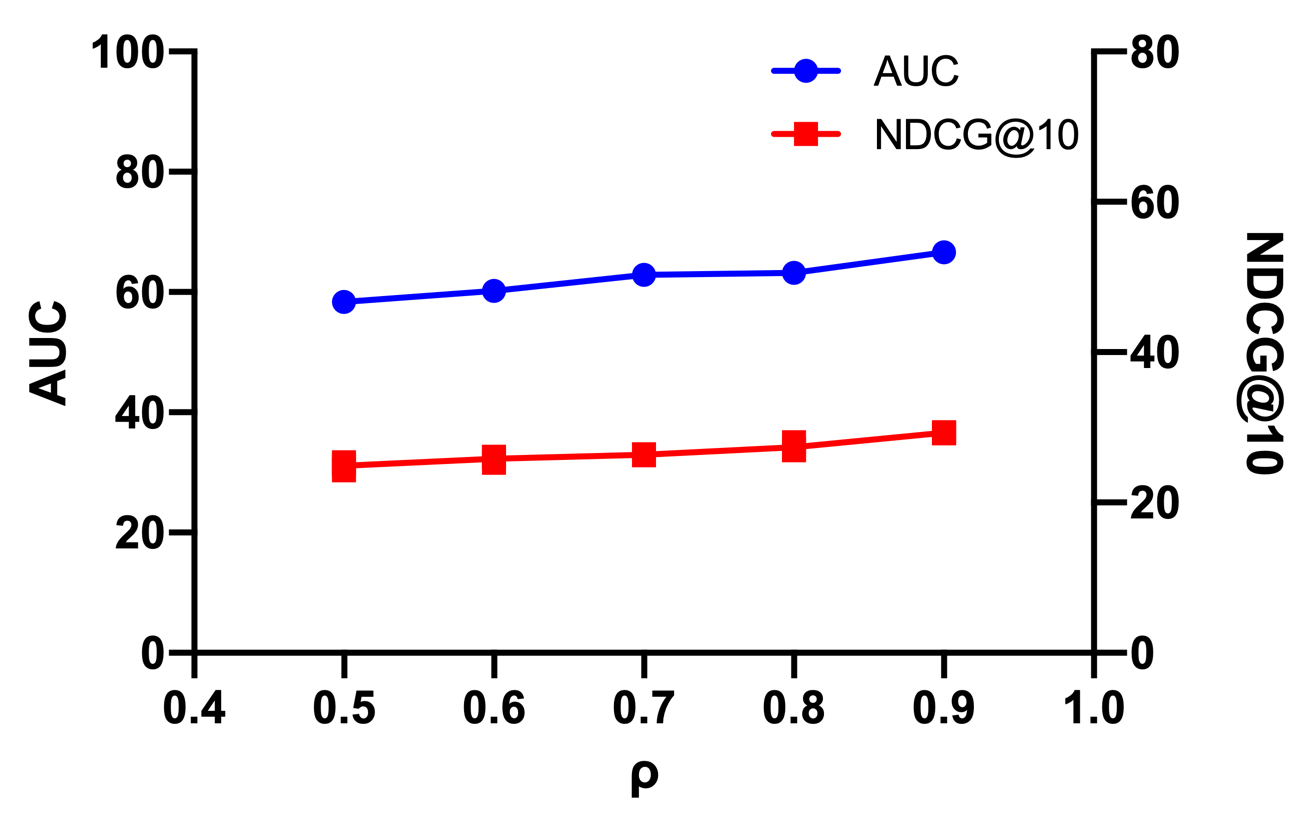}
		\caption{STTFedRec-DSSM}
		\label{rho-a}
	\end{subfigure}
	\begin{subfigure}{0.49\linewidth}
		\centering
		\includegraphics[width=\linewidth]{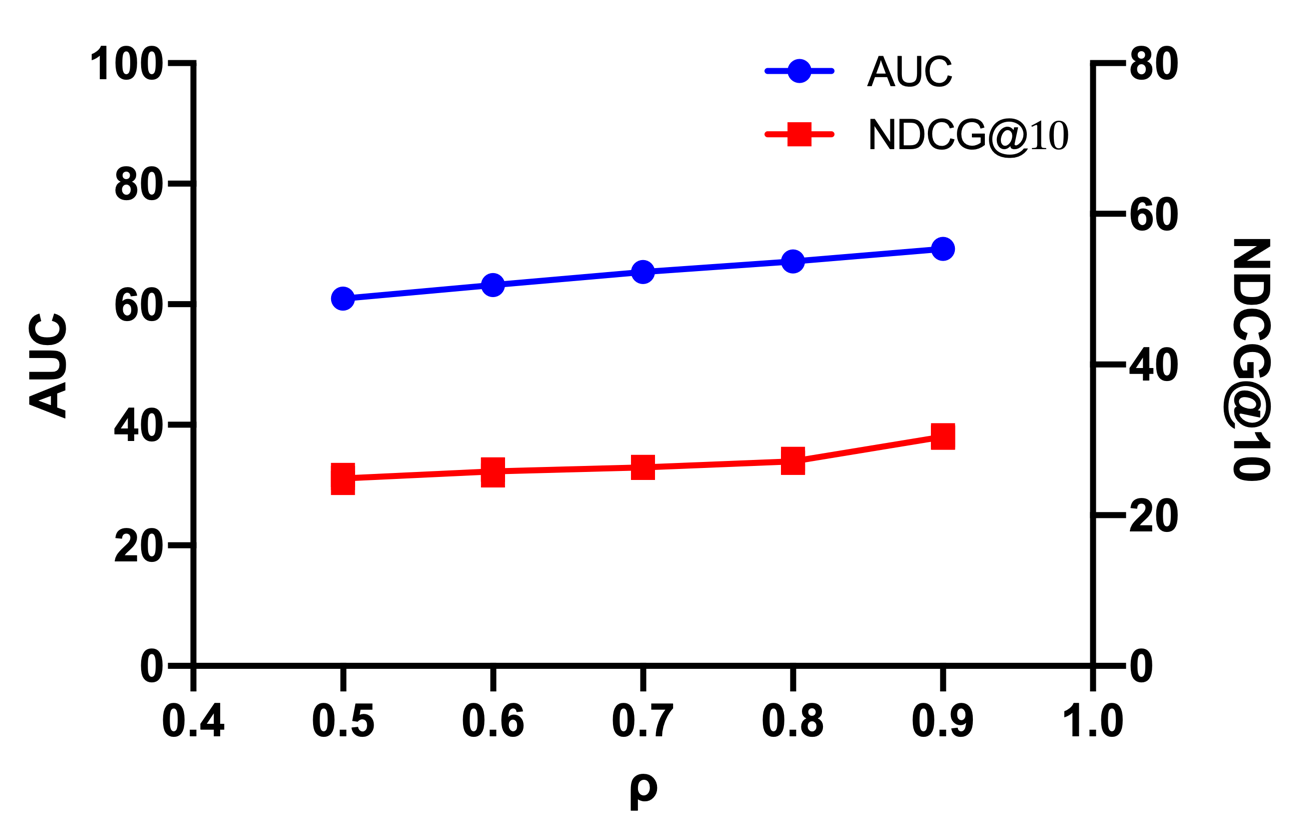}
		\caption{STTFedRec-CLSM}
		\label{rho-b}
	\end{subfigure}
	\caption{The performance by factor of negative sampling rate on Adressa}
	\label{rho}
\end{figure}

\textbf{Effect of $\rho$.} As the results are shown in Figure \ref{rho}, The AUC and NDCG@10 of both STTFedRec-DSSM and STTFedRec-CLSM increased with the increasing negative sampling rate, i.e., the higher the negative sampling rate, the better the prediction performance of STTFedRec. The result consistent with the observation of Shen et al.\cite{shen2014latent} in their experiments for centralized two-tower model

\begin{figure}[h]
    \centering
    \includegraphics[width=\linewidth]{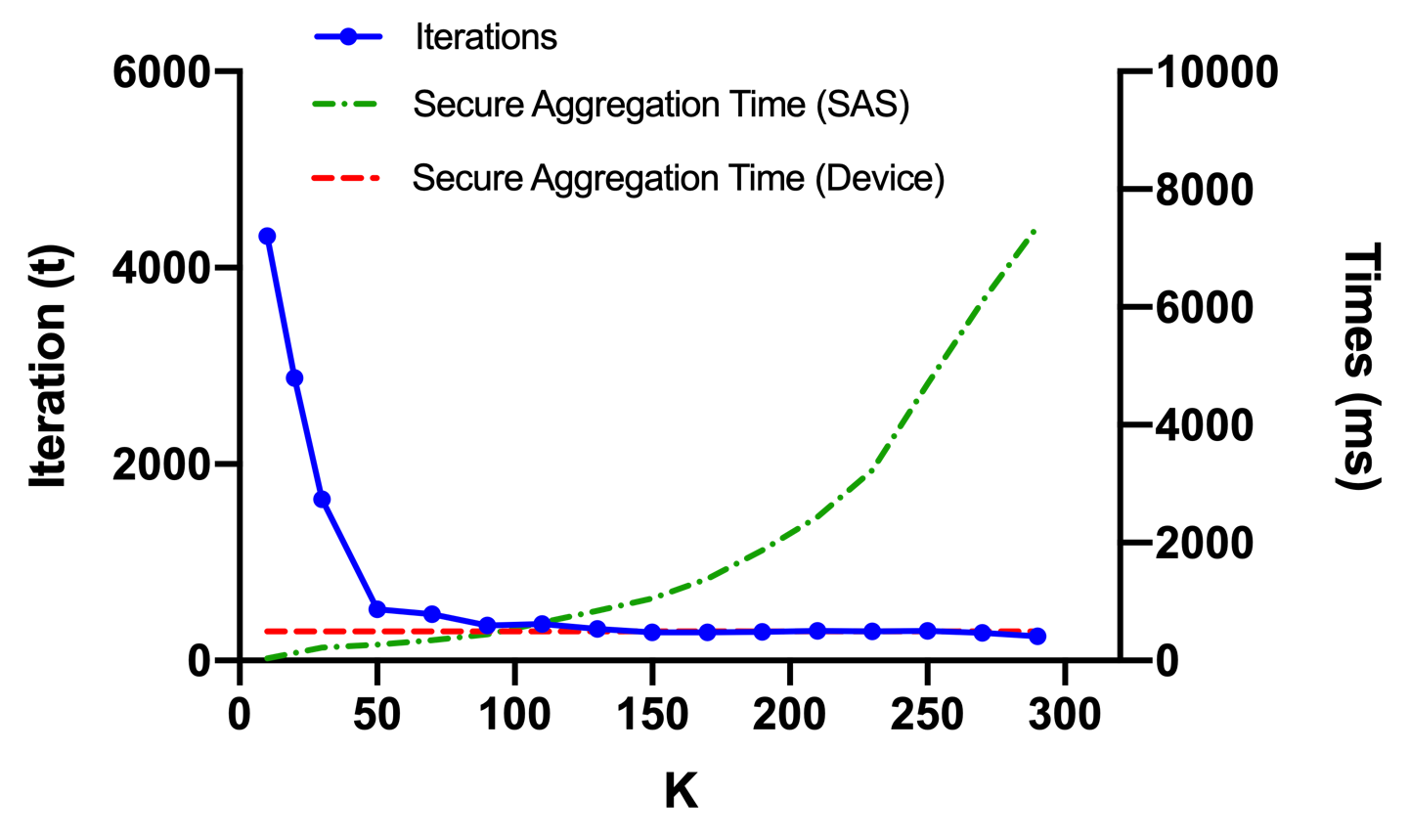}
	\caption{The efficiency by factor of K on Adressa (STTFedRec-DSSM)}
    \label{K-a}
\end{figure}

\begin{figure}[h]
    \centering
    \includegraphics[width=\linewidth]{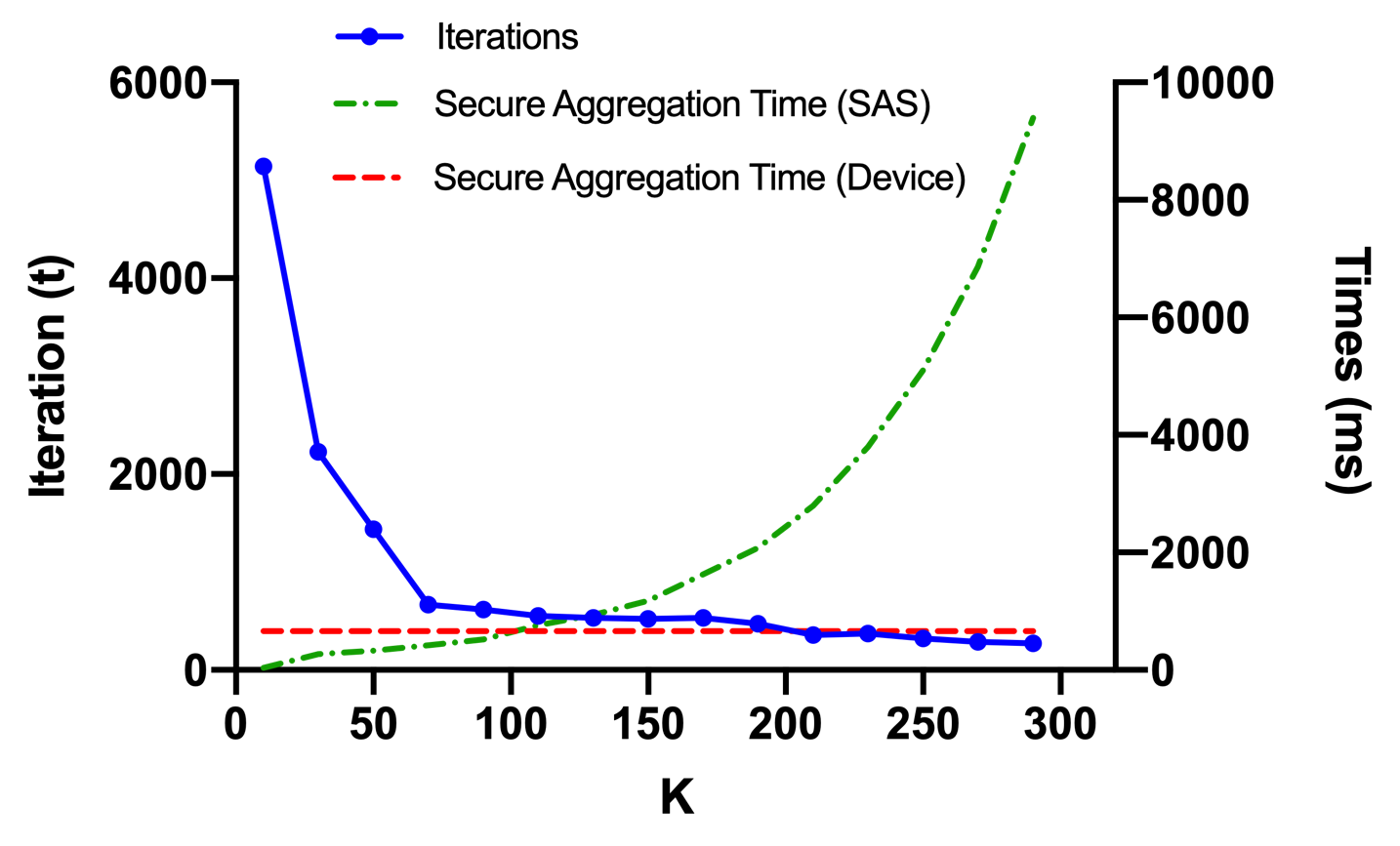}
	\caption{The efficiency by factor of K on Adressa (STTFedRec-CLSM)}
    \label{K-b}
\end{figure}

\textbf{Effect of $K$.} The comparison result with different $K$ as shown in figure \ref{K-a} and figure \ref{K-b}. The selected user size for participating in one round of training affects convergence and secure aggregation efficiency. Specifically, we have the following observations: (1) The larger the number of users participating in a single training round, the more efficient the model convergence. (2) With the increase of $K$, the longer the security aggregation time required by SAS as the number of parameter matrices uploaded to SAS is increasing. (3) The secure aggregation time of the client device is not affected by K since our proposed circular secret sharing chain only requires devices to share secrets with neighboring members. The number of parameter matrices computed does not vary with the total number of members.

\section{Related Work}
Communication and computation can be major bottlenecks in federated learning\cite{kairouz2021advances}, as the end-user Internet connection rate is typically slower than links within servers and can be expensive and unreliable. The processors of mobile devices are also not yet comparable to server-side neural network processors, and user devices are only optimally suited to participate in the training of small-scale deep models.

The current FL based recommendation studies require that the model updates are fully computed in the user device\cite{ammad2019federated}\cite{flanagan2020federated}\cite{chen2020practical}\cite{qi2020privacy}\cite{zhou2019privacy}, which is an unacceptable computational cost for mobile devices with limited resources, especially since the size of recommendation models in the industry is enormous. Muhammad et al.\cite{muhammad2020fedfast} proposed FedFast to accelerate the training efficiency of federation recommendation models. FedFast utilizes the ActvSAMP method to build user clusters from the gradients uploaded at the user, randomly select users in different clusters in each round to participate in federation training and update the local models of similar users by parameter aggregation. This approach can improve the training efficiency globally but cannot reduce the computational effort of the user device. Khan et al.\cite{khan2021payload} introduce the multi-arm bandit mechanism to tackle the item-dependent payloads problem. However, these two methods cannot reduce the device’s workload during the training phase. Qin et al.\cite{qin2021novel} propose a privacy-preserving recommendation system framework based on federated learning(PPRSF). The PPRSF utilizes the privacy hierarchy mechanism, where explicit user feedback records are considered public information, allowing the server to preprocess the model and complete the computationally costly recall module training. PPRSF improves efﬁciency by sacriﬁcing user privacy. 

\section{Conclusion and Future Work}
This paper aims to solve the computation and communication efficiency problem in cross-device FL recommendations. To do this, we introduces split learning into the two-tower recommendation models and proposes STTFedRec, a privacy-preserving and efficient cross-device federated recommendation framework. STTFedRec achieves local computation reduction by splitting the training and computation of the item model from user devices to a performance-powered server. The server with the item model provides low-dimensional item embeddings instead of raw item data to the user devices for local training and online inferring, achieving server broadcast compression. The user devices only need to perform similarity calculations with cached user embeddings to achieve efficient online inferring. We also propose an obfuscated item request strategy and multi-party circular secret sharing chain to enhance the privacy protection of model training. The experiments conducted on two public datasets demonstrate that STTFedRec improves the average computation time and communication size of the baseline models by about 40$\times$ and 42$\times$ in the best case scenario with balanced recommendation accuracy

In the next phase of study, we will further optimize the representation structure of STTFedRec to achieve more competitive prediction performance in specific recommendation scenarios.


\bibliographystyle{ACM-Reference-Format}
\bibliography{ref}


\begin{thebibliography}{42}


\ifx \showCODEN    \undefined \def \showCODEN     #1{\unskip}     \fi
\ifx \showDOI      \undefined \def \showDOI       #1{#1}\fi
\ifx \showISBNx    \undefined \def \showISBNx     #1{\unskip}     \fi
\ifx \showISBNxiii \undefined \def \showISBNxiii  #1{\unskip}     \fi
\ifx \showISSN     \undefined \def \showISSN      #1{\unskip}     \fi
\ifx \showLCCN     \undefined \def \showLCCN      #1{\unskip}     \fi
\ifx \shownote     \undefined \def \shownote      #1{#1}          \fi
\ifx \showarticletitle \undefined \def \showarticletitle #1{#1}   \fi
\ifx \showURL      \undefined \def \showURL       {\relax}        \fi
\providecommand\bibfield[2]{#2}
\providecommand\bibinfo[2]{#2}
\providecommand\natexlab[1]{#1}
\providecommand\showeprint[2][]{arXiv:#2}

\bibitem[Acun et~al\mbox{.}(2021)]%
        {acun2021understanding}
\bibfield{author}{\bibinfo{person}{Bilge Acun}, \bibinfo{person}{Matthew
  Murphy}, \bibinfo{person}{Xiaodong Wang}, \bibinfo{person}{Jade Nie},
  \bibinfo{person}{Carole{-}Jean Wu}, {and} \bibinfo{person}{Kim~M.
  Hazelwood}.} \bibinfo{year}{2021}\natexlab{}.
\newblock \showarticletitle{Understanding Training Efficiency of Deep Learning
  Recommendation Models at Scale}. In \bibinfo{booktitle}{\emph{{IEEE}
  International Symposium on High-Performance Computer Architecture, {HPCA}
  2021, Seoul, South Korea, February 27 - March 3, 2021}}.
  \bibinfo{publisher}{{IEEE}}, \bibinfo{pages}{802--814}.
\newblock


\bibitem[Ammad{-}ud{-}din et~al\mbox{.}(2019)]%
        {ammad2019federated}
\bibfield{author}{\bibinfo{person}{Muhammad Ammad{-}ud{-}din},
  \bibinfo{person}{Elena Ivannikova}, \bibinfo{person}{Suleiman~A. Khan},
  \bibinfo{person}{Were Oyomno}, \bibinfo{person}{Qiang Fu},
  \bibinfo{person}{Kuan~Eeik Tan}, {and} \bibinfo{person}{Adrian Flanagan}.}
  \bibinfo{year}{2019}\natexlab{}.
\newblock \showarticletitle{Federated Collaborative Filtering for
  Privacy-Preserving Personalized Recommendation System}.
\newblock \bibinfo{journal}{\emph{CoRR}}  \bibinfo{volume}{abs/1901.09888}
  (\bibinfo{year}{2019}).
\newblock
\urldef\tempurl%
\url{http://arxiv.org/abs/1901.09888}
\showURL{%
\tempurl}


\bibitem[Anelli et~al\mbox{.}(2019)]%
        {anelli2019towards}
\bibfield{author}{\bibinfo{person}{Vito~Walter Anelli}, \bibinfo{person}{Yashar
  Deldjoo}, \bibinfo{person}{Tommaso~Di Noia}, {and} \bibinfo{person}{Antonio
  Ferrara}.} \bibinfo{year}{2019}\natexlab{}.
\newblock \showarticletitle{Towards Effective Device-Aware Federated Learning}.
  In \bibinfo{booktitle}{\emph{AI*IA 2019 - Advances in Artificial Intelligence
  - XVIIIth International Conference of the Italian Association for Artificial
  Intelligence, Rende, Italy, November 19-22, 2019, Proceedings}}
  \emph{(\bibinfo{series}{Lecture Notes in Computer Science},
  Vol.~\bibinfo{volume}{11946})}. \bibinfo{publisher}{Springer},
  \bibinfo{pages}{477--491}.
\newblock


\bibitem[Chai et~al\mbox{.}(2020)]%
        {chai2020secure}
\bibfield{author}{\bibinfo{person}{Di Chai}, \bibinfo{person}{Leye Wang},
  \bibinfo{person}{Kai Chen}, {and} \bibinfo{person}{Qiang Yang}.}
  \bibinfo{year}{2020}\natexlab{}.
\newblock \showarticletitle{Secure federated matrix factorization}.
\newblock \bibinfo{journal}{\emph{IEEE Intelligent Systems}}
  \bibinfo{volume}{36}, \bibinfo{number}{5} (\bibinfo{year}{2020}),
  \bibinfo{pages}{11--20}.
\newblock


\bibitem[Chen et~al\mbox{.}(2020)]%
        {chen2020practical}
\bibfield{author}{\bibinfo{person}{Chaochao Chen}, \bibinfo{person}{Jun Zhou},
  \bibinfo{person}{Bingzhe Wu}, \bibinfo{person}{Wenjing Fang},
  \bibinfo{person}{Li Wang}, \bibinfo{person}{Yuan Qi}, {and}
  \bibinfo{person}{Xiaolin Zheng}.} \bibinfo{year}{2020}\natexlab{}.
\newblock \showarticletitle{Practical privacy preserving POI recommendation}.
\newblock \bibinfo{journal}{\emph{ACM Transactions on Intelligent Systems and
  Technology (TIST)}} \bibinfo{volume}{11}, \bibinfo{number}{5}
  (\bibinfo{year}{2020}), \bibinfo{pages}{1--20}.
\newblock


\bibitem[Dwork et~al\mbox{.}(2016)]%
        {dwork2016calibrating}
\bibfield{author}{\bibinfo{person}{Cynthia Dwork}, \bibinfo{person}{Frank
  McSherry}, \bibinfo{person}{Kobbi Nissim}, {and} \bibinfo{person}{Adam
  Smith}.} \bibinfo{year}{2016}\natexlab{}.
\newblock \showarticletitle{Calibrating noise to sensitivity in private data
  analysis}.
\newblock \bibinfo{journal}{\emph{Journal of Privacy and Confidentiality}}
  \bibinfo{volume}{7}, \bibinfo{number}{3} (\bibinfo{year}{2016}),
  \bibinfo{pages}{17--51}.
\newblock


\bibitem[Dwork et~al\mbox{.}(2014)]%
        {dwork2014algorithmic}
\bibfield{author}{\bibinfo{person}{Cynthia Dwork}, \bibinfo{person}{Aaron
  Roth}, {et~al\mbox{.}}} \bibinfo{year}{2014}\natexlab{}.
\newblock \showarticletitle{The algorithmic foundations of differential
  privacy.}
\newblock \bibinfo{journal}{\emph{Found. Trends Theor. Comput. Sci.}}
  \bibinfo{volume}{9}, \bibinfo{number}{3-4} (\bibinfo{year}{2014}),
  \bibinfo{pages}{211--407}.
\newblock


\bibitem[Elkahky et~al\mbox{.}(2015)]%
        {elkahky2015multi}
\bibfield{author}{\bibinfo{person}{Ali~Mamdouh Elkahky}, \bibinfo{person}{Yang
  Song}, {and} \bibinfo{person}{Xiaodong He}.} \bibinfo{year}{2015}\natexlab{}.
\newblock \showarticletitle{A Multi-View Deep Learning Approach for Cross
  Domain User Modeling in Recommendation Systems}. In
  \bibinfo{booktitle}{\emph{Proceedings of the 24th International Conference on
  World Wide Web, {WWW} 2015, Florence, Italy, May 18-22, 2015}},
  \bibfield{editor}{\bibinfo{person}{Aldo Gangemi}, \bibinfo{person}{Stefano
  Leonardi}, {and} \bibinfo{person}{Alessandro Panconesi}} (Eds.).
  \bibinfo{publisher}{{ACM}}, \bibinfo{pages}{278--288}.
\newblock


\bibitem[Fang and Qian(2021)]%
        {fang2021privacy}
\bibfield{author}{\bibinfo{person}{Haokun Fang} {and} \bibinfo{person}{Quan
  Qian}.} \bibinfo{year}{2021}\natexlab{}.
\newblock \showarticletitle{Privacy preserving machine learning with
  homomorphic encryption and federated learning}.
\newblock \bibinfo{journal}{\emph{Future Internet}} \bibinfo{volume}{13},
  \bibinfo{number}{4} (\bibinfo{year}{2021}), \bibinfo{pages}{94}.
\newblock


\bibitem[Flanagan et~al\mbox{.}(2020)]%
        {flanagan2020federated}
\bibfield{author}{\bibinfo{person}{Adrian Flanagan}, \bibinfo{person}{Were
  Oyomno}, \bibinfo{person}{Alexander Grigorievskiy},
  \bibinfo{person}{Kuan~Eeik Tan}, \bibinfo{person}{Suleiman~A. Khan}, {and}
  \bibinfo{person}{Muhammad Ammad{-}ud{-}din}.}
  \bibinfo{year}{2020}\natexlab{}.
\newblock \showarticletitle{Federated Multi-view Matrix Factorization for
  Personalized Recommendations}. In \bibinfo{booktitle}{\emph{Machine Learning
  and Knowledge Discovery in Databases - European Conference, {ECML} {PKDD}
  2020, Ghent, Belgium, September 14-18, 2020, Proceedings, Part {II}}}
  \emph{(\bibinfo{series}{Lecture Notes in Computer Science},
  Vol.~\bibinfo{volume}{12458})}, \bibfield{editor}{\bibinfo{person}{Frank
  Hutter}, \bibinfo{person}{Kristian Kersting}, \bibinfo{person}{Jefrey
  Lijffijt}, {and} \bibinfo{person}{Isabel Valera}} (Eds.).
  \bibinfo{publisher}{Springer}, \bibinfo{pages}{324--347}.
\newblock


\bibitem[Gulla et~al\mbox{.}(2017)]%
        {gulla2017adressa}
\bibfield{author}{\bibinfo{person}{Jon~Atle Gulla}, \bibinfo{person}{Lemei
  Zhang}, \bibinfo{person}{Peng Liu}, \bibinfo{person}{{\"{O}}zlem
  {\"{O}}zg{\"{o}}bek}, {and} \bibinfo{person}{Xiaomeng Su}.}
  \bibinfo{year}{2017}\natexlab{}.
\newblock \showarticletitle{The Adressa dataset for news recommendation}. In
  \bibinfo{booktitle}{\emph{Proceedings of the International Conference on Web
  Intelligence, Leipzig, Germany, August 23-26, 2017}},
  \bibfield{editor}{\bibinfo{person}{Amit~P. Sheth}, \bibinfo{person}{Axel
  Ngonga}, \bibinfo{person}{Yin Wang}, \bibinfo{person}{Elizabeth Chang},
  \bibinfo{person}{Dominik Slezak}, \bibinfo{person}{Bogdan Franczyk},
  \bibinfo{person}{Rainer Alt}, \bibinfo{person}{Xiaohui Tao}, {and}
  \bibinfo{person}{Rainer Unland}} (Eds.). \bibinfo{publisher}{{ACM}},
  \bibinfo{pages}{1042--1048}.
\newblock


\bibitem[Gupta and Raskar(2018)]%
        {gupta2018distributed}
\bibfield{author}{\bibinfo{person}{Otkrist Gupta} {and} \bibinfo{person}{Ramesh
  Raskar}.} \bibinfo{year}{2018}\natexlab{}.
\newblock \showarticletitle{Distributed learning of deep neural network over
  multiple agents}.
\newblock \bibinfo{journal}{\emph{Journal of Network and Computer
  Applications}}  \bibinfo{volume}{116} (\bibinfo{year}{2018}),
  \bibinfo{pages}{1--8}.
\newblock


\bibitem[Harper and Konstan(2015)]%
        {harper2015movielens}
\bibfield{author}{\bibinfo{person}{F~Maxwell Harper} {and}
  \bibinfo{person}{Joseph~A Konstan}.} \bibinfo{year}{2015}\natexlab{}.
\newblock \showarticletitle{The movielens datasets: History and context}.
\newblock \bibinfo{journal}{\emph{Acm transactions on interactive intelligent
  systems (tiis)}} \bibinfo{volume}{5}, \bibinfo{number}{4}
  (\bibinfo{year}{2015}), \bibinfo{pages}{1--19}.
\newblock


\bibitem[Hu et~al\mbox{.}(2020)]%
        {hu2020graph}
\bibfield{author}{\bibinfo{person}{Linmei Hu}, \bibinfo{person}{Siyong Xu},
  \bibinfo{person}{Chen Li}, \bibinfo{person}{Cheng Yang},
  \bibinfo{person}{Chuan Shi}, \bibinfo{person}{Nan Duan},
  \bibinfo{person}{Xing Xie}, {and} \bibinfo{person}{Ming Zhou}.}
  \bibinfo{year}{2020}\natexlab{}.
\newblock \showarticletitle{Graph Neural News Recommendation with Unsupervised
  Preference Disentanglement}. In \bibinfo{booktitle}{\emph{Proceedings of the
  58th Annual Meeting of the Association for Computational Linguistics, {ACL}
  2020, Online, July 5-10, 2020}}, \bibfield{editor}{\bibinfo{person}{Dan
  Jurafsky}, \bibinfo{person}{Joyce Chai}, \bibinfo{person}{Natalie Schluter},
  {and} \bibinfo{person}{Joel~R. Tetreault}} (Eds.).
  \bibinfo{publisher}{Association for Computational Linguistics},
  \bibinfo{pages}{4255--4264}.
\newblock


\bibitem[Huang et~al\mbox{.}(2020b)]%
        {huang2020embedding}
\bibfield{author}{\bibinfo{person}{Jui{-}Ting Huang}, \bibinfo{person}{Ashish
  Sharma}, \bibinfo{person}{Shuying Sun}, \bibinfo{person}{Li Xia},
  \bibinfo{person}{David Zhang}, \bibinfo{person}{Philip Pronin},
  \bibinfo{person}{Janani Padmanabhan}, \bibinfo{person}{Giuseppe Ottaviano},
  {and} \bibinfo{person}{Linjun Yang}.} \bibinfo{year}{2020}\natexlab{b}.
\newblock \showarticletitle{Embedding-based Retrieval in Facebook Search}. In
  \bibinfo{booktitle}{\emph{{KDD} '20: The 26th {ACM} {SIGKDD} Conference on
  Knowledge Discovery and Data Mining, Virtual Event, CA, USA, August 23-27,
  2020}}, \bibfield{editor}{\bibinfo{person}{Rajesh Gupta},
  \bibinfo{person}{Yan Liu}, \bibinfo{person}{Jiliang Tang}, {and}
  \bibinfo{person}{B.~Aditya Prakash}} (Eds.). \bibinfo{publisher}{{ACM}},
  \bibinfo{pages}{2553--2561}.
\newblock


\bibitem[Huang et~al\mbox{.}(2020a)]%
        {huang2020federated}
\bibfield{author}{\bibinfo{person}{Mingkai Huang}, \bibinfo{person}{Hao Li},
  \bibinfo{person}{Bing Bai}, \bibinfo{person}{Chang Wang},
  \bibinfo{person}{Kun Bai}, {and} \bibinfo{person}{Fei Wang}.}
  \bibinfo{year}{2020}\natexlab{a}.
\newblock \showarticletitle{A Federated Multi-View Deep Learning Framework for
  Privacy-Preserving Recommendations}.
\newblock \bibinfo{journal}{\emph{CoRR}}  \bibinfo{volume}{abs/2008.10808}
  (\bibinfo{year}{2020}).
\newblock
\showeprint[arXiv]{2008.10808}
\urldef\tempurl%
\url{https://arxiv.org/abs/2008.10808}
\showURL{%
\tempurl}


\bibitem[Huang et~al\mbox{.}(2013)]%
        {huang2013learning}
\bibfield{author}{\bibinfo{person}{Po{-}Sen Huang}, \bibinfo{person}{Xiaodong
  He}, \bibinfo{person}{Jianfeng Gao}, \bibinfo{person}{Li Deng},
  \bibinfo{person}{Alex Acero}, {and} \bibinfo{person}{Larry~P. Heck}.}
  \bibinfo{year}{2013}\natexlab{}.
\newblock \showarticletitle{Learning deep structured semantic models for web
  search using clickthrough data}. In \bibinfo{booktitle}{\emph{22nd {ACM}
  International Conference on Information and Knowledge Management, CIKM'13,
  San Francisco, CA, USA, October 27 - November 1, 2013}},
  \bibfield{editor}{\bibinfo{person}{Qi~He}, \bibinfo{person}{Arun Iyengar},
  \bibinfo{person}{Wolfgang Nejdl}, \bibinfo{person}{Jian Pei}, {and}
  \bibinfo{person}{Rajeev Rastogi}} (Eds.). \bibinfo{publisher}{{ACM}},
  \bibinfo{pages}{2333--2338}.
\newblock


\bibitem[Kairouz et~al\mbox{.}(2021)]%
        {kairouz2021advances}
\bibfield{author}{\bibinfo{person}{Peter Kairouz}, \bibinfo{person}{H~Brendan
  McMahan}, \bibinfo{person}{Brendan Avent}, \bibinfo{person}{Aur{\'e}lien
  Bellet}, \bibinfo{person}{Mehdi Bennis}, \bibinfo{person}{Arjun~Nitin
  Bhagoji}, \bibinfo{person}{Kallista Bonawitz}, \bibinfo{person}{Zachary
  Charles}, \bibinfo{person}{Graham Cormode}, \bibinfo{person}{Rachel
  Cummings}, {et~al\mbox{.}}} \bibinfo{year}{2021}\natexlab{}.
\newblock \showarticletitle{Advances and open problems in federated learning}.
\newblock \bibinfo{journal}{\emph{Foundations and Trends{\textregistered} in
  Machine Learning}} \bibinfo{volume}{14}, \bibinfo{number}{1--2}
  (\bibinfo{year}{2021}), \bibinfo{pages}{1--210}.
\newblock


\bibitem[Khan et~al\mbox{.}(2021)]%
        {khan2021payload}
\bibfield{author}{\bibinfo{person}{Farwa~K. Khan}, \bibinfo{person}{Adrian
  Flanagan}, \bibinfo{person}{Kuan~Eeik Tan}, \bibinfo{person}{Zareen Alamgir},
  {and} \bibinfo{person}{Muhammad Ammad{-}ud{-}din}.}
  \bibinfo{year}{2021}\natexlab{}.
\newblock \showarticletitle{A Payload Optimization Method for Federated
  Recommender Systems}. In \bibinfo{booktitle}{\emph{RecSys '21: Fifteenth
  {ACM} Conference on Recommender Systems, Amsterdam, The Netherlands, 27
  September 2021 - 1 October 2021}},
  \bibfield{editor}{\bibinfo{person}{Humberto Jes{\'{u}}s~Corona
  Pamp{\'{\i}}n}, \bibinfo{person}{Martha~A. Larson},
  \bibinfo{person}{Martijn~C. Willemsen}, \bibinfo{person}{Joseph~A. Konstan},
  \bibinfo{person}{Julian~J. McAuley}, \bibinfo{person}{Jean
  Garcia{-}Gathright}, \bibinfo{person}{Bouke Huurnink}, {and}
  \bibinfo{person}{Even Oldridge}} (Eds.). \bibinfo{publisher}{{ACM}},
  \bibinfo{pages}{432--442}.
\newblock


\bibitem[Li et~al\mbox{.}(2020)]%
        {li2020federated}
\bibfield{author}{\bibinfo{person}{Tian Li}, \bibinfo{person}{Anit~Kumar Sahu},
  \bibinfo{person}{Ameet Talwalkar}, {and} \bibinfo{person}{Virginia Smith}.}
  \bibinfo{year}{2020}\natexlab{}.
\newblock \showarticletitle{Federated learning: Challenges, methods, and future
  directions}.
\newblock \bibinfo{journal}{\emph{IEEE Signal Processing Magazine}}
  \bibinfo{volume}{37}, \bibinfo{number}{3} (\bibinfo{year}{2020}),
  \bibinfo{pages}{50--60}.
\newblock


\bibitem[Lim et~al\mbox{.}(2020)]%
        {lim2020federated}
\bibfield{author}{\bibinfo{person}{Wei Yang~Bryan Lim},
  \bibinfo{person}{Nguyen~Cong Luong}, \bibinfo{person}{Dinh~Thai Hoang},
  \bibinfo{person}{Yutao Jiao}, \bibinfo{person}{Ying-Chang Liang},
  \bibinfo{person}{Qiang Yang}, \bibinfo{person}{Dusit Niyato}, {and}
  \bibinfo{person}{Chunyan Miao}.} \bibinfo{year}{2020}\natexlab{}.
\newblock \showarticletitle{Federated learning in mobile edge networks: A
  comprehensive survey}.
\newblock \bibinfo{journal}{\emph{IEEE Communications Surveys \& Tutorials}}
  \bibinfo{volume}{22}, \bibinfo{number}{3} (\bibinfo{year}{2020}),
  \bibinfo{pages}{2031--2063}.
\newblock


\bibitem[Lindell(2005)]%
        {lindell2005secure}
\bibfield{author}{\bibinfo{person}{Yehida Lindell}.}
  \bibinfo{year}{2005}\natexlab{}.
\newblock \showarticletitle{Secure multiparty computation for privacy
  preserving data mining}.
\newblock In \bibinfo{booktitle}{\emph{Encyclopedia of Data Warehousing and
  Mining}}. \bibinfo{publisher}{IGI global}, \bibinfo{pages}{1005--1009}.
\newblock


\bibitem[McMahan et~al\mbox{.}(2017)]%
        {mcmahan2017communication}
\bibfield{author}{\bibinfo{person}{Brendan McMahan}, \bibinfo{person}{Eider
  Moore}, \bibinfo{person}{Daniel Ramage}, \bibinfo{person}{Seth Hampson},
  {and} \bibinfo{person}{Blaise~Ag{\"{u}}era y Arcas}.}
  \bibinfo{year}{2017}\natexlab{}.
\newblock \showarticletitle{Communication-Efficient Learning of Deep Networks
  from Decentralized Data}. In \bibinfo{booktitle}{\emph{Proceedings of the
  20th International Conference on Artificial Intelligence and Statistics,
  {AISTATS} 2017, 20-22 April 2017, Fort Lauderdale, FL, {USA}}}
  \emph{(\bibinfo{series}{Proceedings of Machine Learning Research},
  Vol.~\bibinfo{volume}{54})}, \bibfield{editor}{\bibinfo{person}{Aarti Singh}
  {and} \bibinfo{person}{Xiaojin~(Jerry) Zhu}} (Eds.).
  \bibinfo{publisher}{{PMLR}}, \bibinfo{pages}{1273--1282}.
\newblock


\bibitem[Mothukuri et~al\mbox{.}(2021)]%
        {mothukuri2021}
\bibfield{author}{\bibinfo{person}{Viraaji Mothukuri}, \bibinfo{person}{Reza~M
  Parizi}, \bibinfo{person}{Seyedamin Pouriyeh}, \bibinfo{person}{Yan Huang},
  \bibinfo{person}{Ali Dehghantanha}, {and} \bibinfo{person}{Gautam
  Srivastava}.} \bibinfo{year}{2021}\natexlab{}.
\newblock \showarticletitle{A survey on security and privacy of federated
  learning}.
\newblock \bibinfo{journal}{\emph{Future Generation Computer Systems}}
  \bibinfo{volume}{115} (\bibinfo{year}{2021}), \bibinfo{pages}{619--640}.
\newblock


\bibitem[Muhammad et~al\mbox{.}(2020)]%
        {muhammad2020fedfast}
\bibfield{author}{\bibinfo{person}{Khalil Muhammad}, \bibinfo{person}{Qinqin
  Wang}, \bibinfo{person}{Diarmuid O'Reilly{-}Morgan},
  \bibinfo{person}{Elias~Z. Tragos}, \bibinfo{person}{Barry Smyth},
  \bibinfo{person}{Neil Hurley}, \bibinfo{person}{James Geraci}, {and}
  \bibinfo{person}{Aonghus Lawlor}.} \bibinfo{year}{2020}\natexlab{}.
\newblock \showarticletitle{FedFast: Going Beyond Average for Faster Training
  of Federated Recommender Systems}. In \bibinfo{booktitle}{\emph{{KDD} '20:
  The 26th {ACM} {SIGKDD} Conference on Knowledge Discovery and Data Mining,
  Virtual Event, CA, USA, August 23-27, 2020}},
  \bibfield{editor}{\bibinfo{person}{Rajesh Gupta}, \bibinfo{person}{Yan Liu},
  \bibinfo{person}{Jiliang Tang}, {and} \bibinfo{person}{B.~Aditya Prakash}}
  (Eds.). \bibinfo{publisher}{{ACM}}, \bibinfo{pages}{1234--1242}.
\newblock


\bibitem[Pulido-Gaytan et~al\mbox{.}(2021)]%
        {pulido2021privacy}
\bibfield{author}{\bibinfo{person}{Bernardo Pulido-Gaytan},
  \bibinfo{person}{Andrei Tchernykh}, \bibinfo{person}{Jorge~M
  Cort{\'e}s-Mendoza}, \bibinfo{person}{Mikhail Babenko}, \bibinfo{person}{Gleb
  Radchenko}, \bibinfo{person}{Arutyun Avetisyan}, {and}
  \bibinfo{person}{Alexander~Yu Drozdov}.} \bibinfo{year}{2021}\natexlab{}.
\newblock \showarticletitle{Privacy-preserving neural networks with Homomorphic
  encryption: Challenges and opportunities}.
\newblock \bibinfo{journal}{\emph{Peer-to-Peer Networking and Applications}}
  \bibinfo{volume}{14}, \bibinfo{number}{3} (\bibinfo{year}{2021}),
  \bibinfo{pages}{1666--1691}.
\newblock


\bibitem[Qi et~al\mbox{.}(2020)]%
        {qi2020privacy}
\bibfield{author}{\bibinfo{person}{Tao Qi}, \bibinfo{person}{Fangzhao Wu},
  \bibinfo{person}{Chuhan Wu}, \bibinfo{person}{Yongfeng Huang}, {and}
  \bibinfo{person}{Xing Xie}.} \bibinfo{year}{2020}\natexlab{}.
\newblock \showarticletitle{Privacy-Preserving News Recommendation Model
  Training via Federated Learning}.
\newblock \bibinfo{journal}{\emph{CoRR}}  \bibinfo{volume}{abs/2003.09592}
  (\bibinfo{year}{2020}).
\newblock
\showeprint[arXiv]{2003.09592}
\urldef\tempurl%
\url{https://arxiv.org/abs/2003.09592}
\showURL{%
\tempurl}


\bibitem[Qin et~al\mbox{.}(2021)]%
        {qin2021novel}
\bibfield{author}{\bibinfo{person}{Jiangcheng Qin}, \bibinfo{person}{Baisong
  Liu}, {and} \bibinfo{person}{Jiangbo Qian}.} \bibinfo{year}{2021}\natexlab{}.
\newblock \showarticletitle{A Novel Privacy-Preserved Recommender System
  Framework Based On Federated Learning}. In \bibinfo{booktitle}{\emph{{ICSIM}
  2021: 2021 The 4th International Conference on Software Engineering and
  Information Management, Yokohama Japan, January 16-18, 2021}},
  \bibfield{editor}{\bibinfo{person}{Yonghui Li} {and} \bibinfo{person}{Hiroaki
  Nishi}} (Eds.). \bibinfo{publisher}{{ACM}}, \bibinfo{pages}{82--88}.
\newblock


\bibitem[Reddi et~al\mbox{.}(2021)]%
        {ReddiCZGRKKM21}
\bibfield{author}{\bibinfo{person}{Sashank~J. Reddi}, \bibinfo{person}{Zachary
  Charles}, \bibinfo{person}{Manzil Zaheer}, \bibinfo{person}{Zachary Garrett},
  \bibinfo{person}{Keith Rush}, \bibinfo{person}{Jakub Kone{\v{c}}n{\'y}},
  \bibinfo{person}{Sanjiv Kumar}, {and} \bibinfo{person}{Hugh~Brendan
  McMahan}.} \bibinfo{year}{2021}\natexlab{}.
\newblock \showarticletitle{Adaptive Federated Optimization}. In
  \bibinfo{booktitle}{\emph{9th International Conference on Learning
  Representations, {ICLR} 2021, Virtual Event, Austria, May 3-7, 2021}}.
  \bibinfo{publisher}{OpenReview.net}.
\newblock


\bibitem[Shen et~al\mbox{.}(2014)]%
        {shen2014latent}
\bibfield{author}{\bibinfo{person}{Yelong Shen}, \bibinfo{person}{Xiaodong He},
  \bibinfo{person}{Jianfeng Gao}, \bibinfo{person}{Li Deng}, {and}
  \bibinfo{person}{Gr{\'{e}}goire Mesnil}.} \bibinfo{year}{2014}\natexlab{}.
\newblock \showarticletitle{A Latent Semantic Model with Convolutional-Pooling
  Structure for Information Retrieval}. In
  \bibinfo{booktitle}{\emph{Proceedings of the 23rd {ACM} International
  Conference on Conference on Information and Knowledge Management, {CIKM}
  2014, Shanghai, China, November 3-7, 2014}},
  \bibfield{editor}{\bibinfo{person}{Jianzhong Li},
  \bibinfo{person}{Xiaoyang~Sean Wang}, \bibinfo{person}{Minos~N. Garofalakis},
  \bibinfo{person}{Ian Soboroff}, \bibinfo{person}{Torsten Suel}, {and}
  \bibinfo{person}{Min Wang}} (Eds.). \bibinfo{publisher}{{ACM}},
  \bibinfo{pages}{101--110}.
\newblock


\bibitem[Singh et~al\mbox{.}(2019)]%
        {singh2019detailed}
\bibfield{author}{\bibinfo{person}{Abhishek Singh}, \bibinfo{person}{Praneeth
  Vepakomma}, \bibinfo{person}{Otkrist Gupta}, {and} \bibinfo{person}{Ramesh
  Raskar}.} \bibinfo{year}{2019}\natexlab{}.
\newblock \showarticletitle{Detailed comparison of communication efficiency of
  split learning and federated learning}.
\newblock \bibinfo{journal}{\emph{CoRR}}  \bibinfo{volume}{abs/1909.09145}
  (\bibinfo{year}{2019}).
\newblock
\showeprint[arXiv]{1909.09145}
\urldef\tempurl%
\url{http://arxiv.org/abs/1909.09145}
\showURL{%
\tempurl}


\bibitem[Sun et~al\mbox{.}(2021)]%
        {sun2021dsmn}
\bibfield{author}{\bibinfo{person}{Weifeng Sun}, \bibinfo{person}{Lijun Zhang},
  \bibinfo{person}{Kangkang Chang}, {and} \bibinfo{person}{Shumiao Yu}.}
  \bibinfo{year}{2021}\natexlab{}.
\newblock \showarticletitle{{DSMN:} {A} Personalized Information Retrieval
  Algorithm Based on Improved {DSSM}}. In
  \bibinfo{booktitle}{\emph{International Joint Conference on Neural Networks,
  {IJCNN} 2021, Shenzhen, China, July 18-22, 2021}}.
  \bibinfo{publisher}{{IEEE}}, \bibinfo{pages}{1--7}.
\newblock


\bibitem[Vepakomma et~al\mbox{.}(2018)]%
        {vepakomma2018split}
\bibfield{author}{\bibinfo{person}{Praneeth Vepakomma},
  \bibinfo{person}{Otkrist Gupta}, \bibinfo{person}{Tristan Swedish}, {and}
  \bibinfo{person}{Ramesh Raskar}.} \bibinfo{year}{2018}\natexlab{}.
\newblock \showarticletitle{Split learning for health: Distributed deep
  learning without sharing raw patient data}.
\newblock \bibinfo{journal}{\emph{CoRR}}  \bibinfo{volume}{abs/1812.00564}
  (\bibinfo{year}{2018}).
\newblock
\showeprint[arXiv]{1812.00564}
\urldef\tempurl%
\url{http://arxiv.org/abs/1812.00564}
\showURL{%
\tempurl}


\bibitem[Wang et~al\mbox{.}(2021b)]%
        {wang2021cross}
\bibfield{author}{\bibinfo{person}{Jinpeng Wang}, \bibinfo{person}{Jieming
  Zhu}, {and} \bibinfo{person}{Xiuqiang He}.} \bibinfo{year}{2021}\natexlab{b}.
\newblock \showarticletitle{Cross-Batch Negative Sampling for Training
  Two-Tower Recommenders}. In \bibinfo{booktitle}{\emph{{SIGIR} '21: The 44th
  International {ACM} {SIGIR} Conference on Research and Development in
  Information Retrieval, Virtual Event, Canada, July 11-15, 2021}},
  \bibfield{editor}{\bibinfo{person}{Fernando Diaz}, \bibinfo{person}{Chirag
  Shah}, \bibinfo{person}{Torsten Suel}, \bibinfo{person}{Pablo Castells},
  \bibinfo{person}{Rosie Jones}, {and} \bibinfo{person}{Tetsuya Sakai}} (Eds.).
  \bibinfo{publisher}{{ACM}}, \bibinfo{pages}{1632--1636}.
\newblock


\bibitem[Wang et~al\mbox{.}(2021a)]%
        {wang2021dcn}
\bibfield{author}{\bibinfo{person}{Ruoxi Wang}, \bibinfo{person}{Rakesh
  Shivanna}, \bibinfo{person}{Derek~Zhiyuan Cheng}, \bibinfo{person}{Sagar
  Jain}, \bibinfo{person}{Dong Lin}, \bibinfo{person}{Lichan Hong}, {and}
  \bibinfo{person}{Ed~H. Chi}.} \bibinfo{year}{2021}\natexlab{a}.
\newblock \showarticletitle{{DCN} {V2:} Improved Deep {\&} Cross Network and
  Practical Lessons for Web-scale Learning to Rank Systems}. In
  \bibinfo{booktitle}{\emph{{WWW} '21: The Web Conference 2021, Virtual Event /
  Ljubljana, Slovenia, April 19-23, 2021}},
  \bibfield{editor}{\bibinfo{person}{Jure Leskovec}, \bibinfo{person}{Marko
  Grobelnik}, \bibinfo{person}{Marc Najork}, \bibinfo{person}{Jie Tang}, {and}
  \bibinfo{person}{Leila Zia}} (Eds.). \bibinfo{publisher}{{ACM} / {IW3C2}},
  \bibinfo{pages}{1785--1797}.
\newblock


\bibitem[Wang et~al\mbox{.}(2020)]%
        {wang2020convergence}
\bibfield{author}{\bibinfo{person}{Xiaofei Wang}, \bibinfo{person}{Yiwen Han},
  \bibinfo{person}{Victor~CM Leung}, \bibinfo{person}{Dusit Niyato},
  \bibinfo{person}{Xueqiang Yan}, {and} \bibinfo{person}{Xu Chen}.}
  \bibinfo{year}{2020}\natexlab{}.
\newblock \showarticletitle{Convergence of edge computing and deep learning: A
  comprehensive survey}.
\newblock \bibinfo{journal}{\emph{IEEE Communications Surveys \& Tutorials}}
  \bibinfo{volume}{22}, \bibinfo{number}{2} (\bibinfo{year}{2020}),
  \bibinfo{pages}{869--904}.
\newblock


\bibitem[Yang et~al\mbox{.}(2021)]%
        {yang2021improvement}
\bibfield{author}{\bibinfo{person}{Fan Yang}, \bibinfo{person}{Huaqiong Wang},
  {and} \bibinfo{person}{Jianjing Fu}.} \bibinfo{year}{2021}\natexlab{}.
\newblock \showarticletitle{Improvement of recommendation algorithm based on
  collaborative deep learning and its parallelization on Spark}.
\newblock \bibinfo{journal}{\emph{J. Parallel and Distrib. Comput.}}
  \bibinfo{volume}{148} (\bibinfo{year}{2021}), \bibinfo{pages}{58--68}.
\newblock


\bibitem[Yang et~al\mbox{.}(2020b)]%
        {yang2020mixed}
\bibfield{author}{\bibinfo{person}{Ji Yang}, \bibinfo{person}{Xinyang Yi},
  \bibinfo{person}{Derek~Zhiyuan Cheng}, \bibinfo{person}{Lichan Hong},
  \bibinfo{person}{Yang Li}, \bibinfo{person}{Simon~Xiaoming Wang},
  \bibinfo{person}{Taibai Xu}, {and} \bibinfo{person}{Ed~H. Chi}.}
  \bibinfo{year}{2020}\natexlab{b}.
\newblock \showarticletitle{Mixed Negative Sampling for Learning Two-tower
  Neural Networks in Recommendations}. In \bibinfo{booktitle}{\emph{Companion
  of The 2020 Web Conference 2020, Taipei, Taiwan, April 20-24, 2020}},
  \bibfield{editor}{\bibinfo{person}{Amal El~Fallah Seghrouchni},
  \bibinfo{person}{Gita Sukthankar}, \bibinfo{person}{Tie{-}Yan Liu}, {and}
  \bibinfo{person}{Maarten van Steen}} (Eds.). \bibinfo{publisher}{{ACM} /
  {IW3C2}}, \bibinfo{pages}{441--447}.
\newblock


\bibitem[Yang et~al\mbox{.}(2020a)]%
        {yang2020federated}
\bibfield{author}{\bibinfo{person}{Liu Yang}, \bibinfo{person}{Ben Tan},
  \bibinfo{person}{Vincent~W. Zheng}, \bibinfo{person}{Kai Chen}, {and}
  \bibinfo{person}{Qiang Yang}.} \bibinfo{year}{2020}\natexlab{a}.
\newblock \showarticletitle{Federated Recommendation Systems}.
\newblock In \bibinfo{booktitle}{\emph{Federated Learning - Privacy and
  Incentive}}, \bibfield{editor}{\bibinfo{person}{Qiang Yang},
  \bibinfo{person}{Lixin Fan}, {and} \bibinfo{person}{Han Yu}} (Eds.).
  \bibinfo{series}{Lecture Notes in Computer Science},
  Vol.~\bibinfo{volume}{12500}. \bibinfo{publisher}{Springer},
  \bibinfo{pages}{225--239}.
\newblock


\bibitem[Yi et~al\mbox{.}(2019)]%
        {yi2019sampling}
\bibfield{author}{\bibinfo{person}{Xinyang Yi}, \bibinfo{person}{Ji Yang},
  \bibinfo{person}{Lichan Hong}, \bibinfo{person}{Derek~Zhiyuan Cheng},
  \bibinfo{person}{Lukasz Heldt}, \bibinfo{person}{Aditee Kumthekar},
  \bibinfo{person}{Zhe Zhao}, \bibinfo{person}{Li Wei}, {and}
  \bibinfo{person}{Ed~H. Chi}.} \bibinfo{year}{2019}\natexlab{}.
\newblock \showarticletitle{Sampling-bias-corrected neural modeling for large
  corpus item recommendations}. In \bibinfo{booktitle}{\emph{Proceedings of the
  13th {ACM} Conference on Recommender Systems, RecSys 2019, Copenhagen,
  Denmark, September 16-20, 2019}}, \bibfield{editor}{\bibinfo{person}{Toine
  Bogers}, \bibinfo{person}{Alan Said}, \bibinfo{person}{Peter Brusilovsky},
  {and} \bibinfo{person}{Domonkos Tikk}} (Eds.). \bibinfo{publisher}{{ACM}},
  \bibinfo{pages}{269--277}.
\newblock


\bibitem[Zhai et~al\mbox{.}(2016)]%
        {zhai2016deepintent}
\bibfield{author}{\bibinfo{person}{Shuangfei Zhai}, \bibinfo{person}{Keng{-}hao
  Chang}, \bibinfo{person}{Ruofei Zhang}, {and}
  \bibinfo{person}{Zhongfei~(Mark) Zhang}.} \bibinfo{year}{2016}\natexlab{}.
\newblock \showarticletitle{DeepIntent: Learning Attentions for Online
  Advertising with Recurrent Neural Networks}. In
  \bibinfo{booktitle}{\emph{Proceedings of the 22nd {ACM} {SIGKDD}
  International Conference on Knowledge Discovery and Data Mining, San
  Francisco, CA, USA, August 13-17, 2016}},
  \bibfield{editor}{\bibinfo{person}{Balaji Krishnapuram},
  \bibinfo{person}{Mohak Shah}, \bibinfo{person}{Alexander~J. Smola},
  \bibinfo{person}{Charu~C. Aggarwal}, \bibinfo{person}{Dou Shen}, {and}
  \bibinfo{person}{Rajeev Rastogi}} (Eds.). \bibinfo{publisher}{{ACM}},
  \bibinfo{pages}{1295--1304}.
\newblock


\bibitem[Zhou et~al\mbox{.}(2019)]%
        {zhou2019privacy}
\bibfield{author}{\bibinfo{person}{Pan Zhou}, \bibinfo{person}{Kehao Wang},
  \bibinfo{person}{Linke Guo}, \bibinfo{person}{Shimin Gong}, {and}
  \bibinfo{person}{Bolong Zheng}.} \bibinfo{year}{2019}\natexlab{}.
\newblock \showarticletitle{A privacy-preserving distributed contextual
  federated online learning framework with big data support in social
  recommender systems}.
\newblock \bibinfo{journal}{\emph{IEEE Transactions on Knowledge and Data
  Engineering}} \bibinfo{volume}{33}, \bibinfo{number}{3}
  (\bibinfo{year}{2019}), \bibinfo{pages}{824--838}.
\newblock


\end{thebibliography}


\end{document}